\documentclass[referee]{aa} % for a paper on 1 column  
\usepackage[bottom]{footmisc}
\usepackage{textgreek}
\usepackage{stfloats}
\usepackage{graphicx}
\usepackage{array}
\usepackage{txfonts}
\begin{document}

   \title{ In situ characterization of volatile and refractory hydrocarbons produced by UV photolysis of interstellar C$_2$H$_2$ ice}

   \subtitle{}

  \titlerunning{UV photo-processing of pure acetylene ice}
\authorrunning{P. Samarth et al.}
   \author{P. Samarth\inst{1,2},
            G. Fedoseev\inst{3},
            M. Bulak\inst{1}, 
            S. Ioppolo\inst{4},
            L. Hornekær\inst{4},
            E. F. van Dishoeck\inst{2},
            H. Linnartz\inst{1}\thanks{Deceased; we acknowledge his contributions to this work.},
            and
            K. -J. Chuang\inst{1}
            }

\institute{
Laboratory for Astrophysics, Leiden Observatory, Leiden University,
P.O. Box 9513, 2300 RA Leiden, The Netherlands\\
\email{samarth@strw.leidenuniv.nl}
\and
Leiden Observatory, Leiden University, P.O. Box 9513,
2300 RA Leiden, The Netherlands
\and
Xinjiang Astronomical Observatory, Chinese Academy of Sciences,
Urumqi 830011, China
\and
Center for Interstellar Catalysis, Department of Physics and Astronomy,
Aarhus University, Aarhus, Denmark, 8000
}

\date{Received: 30 July 2025; accepted 30 November 2025}
 
  \abstract
   {Acetylene (C$_2$H$_2$) has been commonly observed in various astronomical objects, including star-forming regions, young stellar objects, and our Solar System. Theoretical and laboratory studies have proposed multiple mechanisms that link this simplest alkyne to volatile hydrocarbons and polycyclic aromatic hydrocarbons (PAHs) through UV- or cosmic-ray-induced energetic processes. However, it is still unclear whether refractory material can be efficiently formed through solid-state reactions involving C$_2$H$_2$ on dust grains.}
   {In this work, we aim to experimentally study the chemical complexity induced by the UV irradiation of pure C$_2$H$_2$ ice and characterize both volatile and nonvolatile photoproducts to better understand the evolution of simple hydrocarbons under astronomically relevant conditions.}
   {Experiments were performed using MATRI$^2$CES, an ultra-high vacuum, cryogenic setup to investigate the C$_2$H$_2$ ice chemistry induced by UV photons between 7.2 to 10.2~eV at 15~K. The UV-processed ice samples were monitored in situ by laser desorption post-ionization reflection time-of-flight mass spectrometry (LDPI ReTOF-MS) in combination with the pulsed ion deflection (PID) technique. The mass spectrometric data of volatiles and refractory residues produced upon VUV photolysis of C$_2$H$_2$ ice were collected in situ at 15~K and 300~K, respectively, minimizing uncertainties associated with external analytical methods used in previous studies.}
   {The experimental results obtained after photolysis of pure C$_2$H$_2$ ice with a fluence of 3 $\times$ 10$^{17}$ photons cm$^{-2}$ (10$^6$ years in dense clouds, show the formation of large saturated and unsaturated hydrocarbons containing up to 13 carbon atoms, including molecules identified in previous similar studies. After the sublimation of these volatile products, measurements of the residue at 300~K revealed a rich and distinct mass spectrum suggesting the synthesis of refractories composed of conjugated triple bonds (-C$\equiv$C-) and double (-C=C-) bonds. The astrochemical implications and the possible connection of the produced residues with unidentified infrared emission bands are discussed.}
 {}
   \keywords{Astrochemistry --
            Methods: laboratory: solid state --
            ISM: molecules --
            Solid state: refractory --
            Solid state: volatile --
            Molecular processes}

    \maketitle 

%
%-------------------------------------------------------------------

\section{\label{sec:level1} Introduction\protect}

Interstellar dust grains are ubiquitous throughout the interstellar medium (ISM) and play a crucial role in the processes of stellar evolution and astrochemistry \citep{Draine_2003}. In addition to blocking out the interstellar radiation field (ISRF), which results in cooler core temperatures (as low as 10~K) within interstellar clouds, dust grains also provide a third body that enables gas phase species to accrete and react via both energetic and non-energetic means \citep{Tielens_Hagen_1982, Hasegawa_Herbst_Leung_1992, Linnartz_Ioppolo_Fedoseev_2015, Cuppen_Walsh_Lamberts_Semenov_Garrod_Penteado_Ioppolo_2017}. This has been demonstrated by the efficient formation of H$_2$ proposed to occur on dust grains at low temperatures \citep{Gould_Salpeter_1963}.

As molecular clouds evolve, dust grains accumulate icy mantles through the accretion of gas phase species and surface reactions on these cold grains drive the formation of both simple and complex molecular ices. These mantles are primarily composed of water (H$_2$O), carbon monoxide (CO), ammonia (NH$_3$), and organic molecules, such as methane (CH$_4$) and methanol (CH$_3$OH); readers can refer to \citet{Boogert_Gerakines_Whittet_2015} for a review. This stage of interstellar icy dust grain evolution is associated with the bottom-up formation of various hydrocarbons and complex organic molecules (COMs) via solid-state chemistry involving both non-energetic reactions, such as those driven by neutral hydrogen bombardment, energetic processes induced by secondary UV photons, and cosmic ray irradiation \citep{Oberg_Garrod_Dishoeck_Linnartz_2009, Modica_Palumbo_2010, Butscher_Duvernay_Danger_Chiavassa_2016, Oberg_2016, Chuang_Fedoseev_Qasim_Ioppolo_vanDishoeck_Linnartz_2017, Fedoseev_Cuppen_Ioppolo_Lamberts_Linnartz_2015, Fedoseev_Qasim_Chuang_Ioppolo_Lamberts_vanDishoeck_Linnartz_2022, Cuppen_Linnartz_Ioppolo_2024}.
 
Additionally, observational evidence indicates that dust grains in diffuse clouds are already coated with refractory organic residues \citep{Sandford_Allamandola_Tielens_Sellgren_Tapia_Pendleton_1991}. These residues are characterized by the aliphatic C–H stretching mode at 3.4 $\mu$m and by weaker absorption features at 6.8 and 7.3 $\mu$m, corresponding to the bending modes of –CH$_3$ and –CH$_2$ groups \citep{Greenberg_Shen_1999, Jones_Fanciullo_KOhler_Verstraete_Guillet_Bocchio_Ysard_2013, Pendleton_Allamandola_2002, Dartois_Caro_Deboffle_d’Hendecourt_2004, Li_Draine_2012}. Evidence for the presence of organic residues is also supported by observations of comets and analyses of carbonaceous chondrites, which are thought to preserve unaltered material from the early stages of stellar evolution \citep{Greenberg_Shalabiea_1994, Pizzarello_Shock_2010}. In the coma of comet 67P/Churyumov-Gerasimenko, hydrocarbons such as hexane (C$_6$H$_{14}$), heptane (C$_7$H$_{16}$), benzene (C$_6$H$_6$), and toluene (C$_7$H$_8$) have been detected by the ROSINA instrument on board the Rosetta space probe, with measurements alluding to the possible presence of even larger nonvolatile aliphatic and aromatic molecules \citep{Schuhmann_Altwegg_Balsiger_Berthelier_DeKeyser_Fiethe_Fuselier_Gasc_Gombosi_Hänni_etal._2019, Fedoseev_Li_Baratta_Palumbo_Chuang_2024}. Additionally, solvent extracts of the carbonaceous asteroid Ryugu reveal the presence of PAHs and refractory organics rich in N, O, and S, along with various other hydrocarbons \citep{Aponte_Dworkin_Glavin_Elsila_Parker_McLain_Naraoka_Okazaki_Takano_Tachibana_etal._2023, Schmitt-Kopplin_Hertkorn_Harir_Moritz_Lucio_Bonal_Quirico_Takano_Dworkin_Naraoka_etal._2023, Yabuta_Cody_Engrand_Kebukawa_DeGregorio_Bonal_Remusat_Stroud_Quirico_Nittler_etal._2023}. 

These organic residues may form inner refractory mantles on dust grains, beneath the ice mantles that accumulate later in molecular clouds. The organic coating on dust grains can significantly modify their chemical and physical properties \citep{Urso_Vuitton_Danger_d’Hendecourt_Flandinet_Djouadi_Mivumbi_Orthous-Daunay_Ruf_Vinogradoff_etal._2020}. This has been demonstrated in models showing that an increase in the sticking threshold of coated silicate grains could promote the growth of larger planetesimals. Additionally, such organic residues have been suggested to enhance the tensile strength of micro-granular pebbles \citep{Kouchi_Kudo_Nakano_Arakawa_Watanabe_Sirono_Higa_Maeno_2002, Bischoff_Kreuzig_Haack_Gundlach_Blum_2020}. 

These inner refractory residues have been suggested to form through prolonged photolysis or radiolysis of volatile hydrocarbons and other carbon-bearing species that accrete onto the surfaces of these dust grains. This can occur in dense molecular clouds where these grains with residues can be recycled to the diffuse clouds \citep{Pendleton_Allamandola_2002, Caro_Schutte_2003}, or within the envelopes of carbon-rich asymptotic giant branch (AGB) stars, where produced gas phase hydrocarbons can condense and undergo energetic processing on dust grain surfaces before being ejected into the ISM \citep{Ehrenfreund_Charnley_2000, Kwok_2004, Kwok_Zhang_2011, VandeSande_Walsh_Millar_2021}.

Within the dust-forming regions of carbon-rich AGB stars, acetylene (C$_2$H$_2$) has been observed to be one of the most abundant gas phase species, with abundances up to $8\times10^{-5}$ relative to H$_2$ \citep{Fonfria_Cernicharo_Richter_Lacy_2008}. This abundance is one to two orders of magnitude higher than that of other carbon-bearing species, consistent with astrochemical models' equilibrium predictions \citep{Agundez_Marcelino_Cabezas_Fuentetaja_Tercero_de_Vicente_Cernicharo_2022}. In diffuse clouds and photodissociation regions (PDRs), C$_2$H$_2$ is thought to form through the dissociation of larger hydrocarbons or PAHs \citep{Zhen_Paardekooper_Candian_Linnartz_Tielens_2014, West_Castillo_Sit_Mohamad_Lowe_Joblin_Bodi_Mayer_2018, Rapacioli_Cazaux_Foley_Simon_Hoekstra_SchlathOlter_2018}. C$_2$H$_2$ has been observed in various astronomical regions in the infrared, such as around massive young stellar objects (YSOs) \citep{Lahuis_van_Dishoeck_2000, Gelder_Francis_Dishoeck_Tychoniec_Ray_Beuther_Garatti_Chen_Devaraj_Gieser_etal._2024}, around carbon stars \citep{Loon_Marshall_Cohen_Matsuura_Wood_Yamamura_Zijlstra_2006}, and in disks around low-mass stars \citep{Pascucci_Apai_Luhman_Henning_Bouwman_Meyer_Lahuis_Natta_2009, Salyk_Pontoppidan_Blake_Najita_Carr_2011}. Recently, C$_2$H$_2$ was detected in the inner regions of many more protoplanetary disks using the JWST Mid-Infrared Instrument (MIRI)
\citep{Dishoeck_Grant_Tabone_Gelder_Francis_Tychoniec_Bettoni_Arabhavi_Gasman_Nazari_etal_2023, Tabone_Bettoni_vanDishoeck_Arabhavi_Grant_Gasman_Henning_Kamp_Güdel_Lagage_etal._2023, Arabhavi_Kamp_Henning_vaDishoeck_Christiaens_Gasman_Perrin_Güdel_Tabone_Kanwar_etal._2024, Colmenares_Bergin_Salyk_Pontoppidan_Arulanantham_Calahan_Banzatti_Andrews_Blake_Ciesla_etal._2024}. C$_2$H$_2$ has also been detected within our Solar System, in the atmospheres of gas giants \citep{Ridgway_1974, Orton_Aitken_Smith_Roche_Caldwell_Snyder_1987}, Saturn's largest moon Titan and trans-Neptunian objects (TNOs) such as Pluto \citep{Kunde_Aikin_Hanel_Jennings_Maguire_Samuelson_1981, Gladstone_Stern_Ennico_Olkin_Weaver_Young_Summers_Strobel_Hinson_Kammer_et_al._2016}. It has been further detected around comets Hyakutake, Halley, and 67P/Churyumov-Gerasimenko \citep{Brooke_Tokunaga_Weaver_Crovisier_Bockelée-Morvan_Crisp_1996, Mumma_DiSanti_Dello_Russo_Magee-Sauer_Gibb_Novak_2003, Roy_Altwegg_Balsiger_Berthelier_Bieler_Briois_Calmonte_Combi_Keyser_Dhooghe_et_al._2015}.

Although C$_2$H$_2$ is widely detected in the gas phase, its solid-state detection remains elusive \citep{Boudin_Schutte_Greenberg_1998}. The IR absorption features of C$_2$H$_2$ ice (at 13.7 $\mu$m) overlap with the stronger H$_2$O ice features between 12 and 16 $\mu$m, making secure identification challenging. Laboratory IR spectroscopic data suggest that C$_2$H$_2$ can be reliably detected only when its concentration relative to H$_2$O exceeds 20\% \citep{Knez_Moore_Ferrante_Hudson_2012}. Nevertheless, observational and theoretical studies suggest that C$_2$H$_2$ ice may exist in molecular clouds. For example, measurements around massive YSOs indicate that the C$_2$H$_2$/H$_2$ ratio increases from 10$^{-8}$–10$^{-9}$ in cooler regions to ~10$^{-6}$ in warmer lines of sight, suggesting possible thermal desorption of C$_2$H$_2$ ices \citep{Lahuis_van_Dishoeck_2000, Sonnentrucker_González-Alfonso_Neufeld_2007}. Although this increase in column density can be partly attributed to the efficient gas phase formation of C$_2$H$_2$ in high temperature (400-700~K) regions, which requires a high activation energy of ~12,000~K (~100 kJ/mol), the gas phase formation of C$_2$H$_2$ is consequently not as efficient in colder regions (10-300~K) \citep{Bast_Lahuis_VanDishoeck_Tielens_2013}. Furthermore, C$_2$H$_2$ has been shown to form in the solid state upon VUV irradiation of methane (CH$_4$) or ethane (C$_2$H$_6$) ices, alluding to the presence of some C$_2$H$_2$ in the solid state \citep{Lo_Lin_Peng_Chou_Lu_Cheng_Ogilvie_2015, Paardekooper_Bossa_Linnartz_2016, Bulak_Paardekooper_Fedoseev_Linnartz_2020, Bennett_Jamieson_Osamura_Kaiser_2006, Ryazantsev_Zasimov_Feldman_2018, Carrascosa2020}. In planetary icy bodies, the C$_2$H$_2$ ice abundance is found to be about 0.1–1\% relative to water \citep{Urso_Vuitton_Danger_d’Hendecourt_Flandinet_Djouadi_Mivumbi_Orthous-Daunay_Ruf_Vinogradoff_etal._2020, Hudson_Moore_1999, Rubin_Altwegg_Balsiger_Berthelier_Combi_DeKeyser_Drozdovskaya_Fiethe_Fuselier_Gasc_etal._2019, Altwegg_Balsiger_Fuselier_2019}. 

In addition to spectral detection constraints, the low abundance of C$_2$H$_2$ ice can be explained by its high chemical reactivity. Over the past two decades, C$_2$H$_2$ has been proposed as a key molecule leading to molecular complexity \citep{Ahrens_Bachmann_Baum_Griesheimer_Kovacs_Weilmuenster_Homann_1994, Ehrenfreund_B_Sephton_2006, Oremland_Voytek_2008, Contreras_Salama_2013, Tielens_2013, Dhanoa_Rawlings_2014, Santoro_Martínez_Lauwaet_Accolla_Tajuelo-Castilla_Merino_Sobrado_Peláez_Herrero_Tanarro_etal._2020, Lo2020,Rap_Schrauwen_Marimuthu_Redlich_Brünken_2022, Chuang_Jaeger_Santos_Henning_2024, Pentsak_Murga_Ananikov_2024}. Experiments have shown that gas phase C$_2$H$_2$ may be crucial in forming larger PAHs via the hydrogen abstraction–acetylene addition (HACA) mechanism \citep{Yang_Kaiser_Troy_Xu_Kostko_Ahmed_Mebel_Zagidullin_Azyazov_2017, Zhao_Kaiser_Xu_Ablikim_Ahmed_Joshi_Veber_Fischer_Mebel_2018, Reizer_Viskolcz_Fiser_2022}. Similarly, HACA has also been shown in the solid state, when energetically processed C$_2$H$_2$ ices resulted in the formation of PAHs such as chrysene (C$_{18}$H$_{12}$) and coronene (C$_{24}$H$_{12}$) \citep{Kaiser_Roessler_1997}. Apart from PAHs, UV photolysis of C$_2$H$_2$ molecules isolated in a neon matrix at 10~K results in the formation of linear carbon chain molecules such as C$_2$H, C$_2$H$_3$, C$_3$, C$_4$, C$_4$H, C$_4$H$_2$, C$_6$, C$_8$, and C$_8$H$_2$ \citep{Wu_Cheng_2008}. Finally, \citet{Cuylle_Zhao_Strazzulla_Linnartz_2014} have reported the formation of polyynes (H-(C$\equiv$C)$_n$-H) up to C$_{20}$H$_2$ upon long-term VUV irradiation of C$_2$H$_2$ ices.

An extensive laboratory study by \citet{Abplanalp_Kaiser_2020} investigated the chemical complexity resulting from 5 keV electron bombardment of pure C$_2$H$_2$ ice at 5~K and reported the formation of multiple hydrocarbon products belonging to the following chemical groups: C$_n$H$_{2n+2}$, C$_n$H$_{2n}$, C$_n$H$_{2n-2}$, C$_n$H$_{2n-4}$, C$_n$H$_{2n-6}$, C$_n$H$_{2n-8}$, C$_n$H$_{2n-10}$, C$_n$H$_{2n-12}$, C$_n$H$_{2n-14}$, and C$_n$H$_{2n-16}$ with \(n \leq 17 \), along with several PAHs. These product assignments were done using single-photon ionization time-of-flight mass spectrometry (SPI ReTOF-MS) and resonance-enhanced multiphoton ionization reflectron time-of-flight mass spectrometry (REMPI ReTOF-MS) during temperature-programmed desorption (TPD) of the photoproducts until 300~K. 

However, a systematic investigation of energetically processed pure C$_2$H$_2$ ice, which can characterize newly formed photoproducts, beyond the 300~K substrate temperature limitations, especially those that desorb above 300~K, such as the nonvolatile residues, remains unavailable. This work presents a series of systematic experiments designed to explore the chemical complexity induced by VUV irradiation of pure C$_2$H$_2$ ices under physical conditions representative of cold, dense regions in the interstellar medium, such as molecular clouds and circumstellar envelopes of carbon-rich AGB stars. This study presents, for the first time, the mass spectra of both volatile photoproducts and the nonvolatile residues at 15~K and 300~K, respectively. The obtained results enable the study of the most pristine organic compounds that originate directly through the UV photolysis of C$_2$H$_2$, preserving the pristine chemical state of the residue, free from alterations by exposure to the atmosphere that typically occur during ex situ sample handling and preparation for en situ analyses. The article is organized as follows. Section~2 describes the experimental setup and the newly implemented high-sensitivity measurement approach. Section~3 presents the results and provides a discussion of their significance. Finally, Section~4 outlines the astronomical implications of the findings.

\section{\label{sec:level2} Experimental methodology \protect}

All experiments were carried out using a custom-made experimental setup, the mass analytical tool to research interstellar ices (MATRI$^2$CES). The setup utilizes a well-optimized localized laser desorption post-ionization time-of-flight mass spectrometry (LDPI ReTOF-MS) to probe the VUV-irradiated C$_2$H$_2$ ice sample. The setup has been introduced and detailed in previous studies \citep{Paardekooper_Bossa_Isokoski_Linnartz_2014, Samarth_Bulak_Paardekooper_Chuang_Linnartz_2024}. The description of the setup and applied experimental methodology relevant to this work is provided in the following section. 

\subsection{\label{sec:level21}Setup description}
MATRI$^2$CES is a cryogenic, ultra-high vacuum (UHV) experimental setup with a base pressure of the mid 10$^{-10}$~mbar, comprising two chambers, the main chamber and the ToF chamber separated by a UHV gate valve. Within the main chamber, the C$_2$H$_2$ gas is deposited onto a gold-plated copper substrate at an incident angle of 5 degrees through a metal capillary tube using a pre-calibrated, high-precision leak valve. The substrate is mounted on the cold finger of a closed-cycle helium cryostat, which sits on a UHV X--Y manipulator. The temperature of the substrate can be regulated in the range from 15 to 300~K using resistive heating coils. 

For our experiments, 100~monolayers (ML) of pure C$_2$H$_2$ ice were deposited on the substrate; 1 ML of ice is assumed to have a column density of 10$^{15}$ molecules cm$^{-2}$. The ice thickness is calculated from the ice deposition rate, which is estimated through laser interferometry using a frequency-stabilized He-Ne laser (632.8 nm) positioned at an angle of 2 degrees to the normal of the substrate. The reflected laser intensity is measured by a photodiode over time as the ice film grows. The absolute ice thickness (\textit{d}) is calculated from the number of fringes observed in the interference pattern \citep{Hudgins_Sandford_Allamandola_Tielens_1993} using the equation

\begin{equation}
d = \frac{m\lambda}{2n_{\text{ice}} \cos\theta_{\text{ice}}},
\label{eq:equation_label01}
\end{equation}

where \(m\) is the number of fringes, \textit{\(\lambda\)} is the wavelength of the laser, \(n_{\text{ice}}\) is the refractive index of the C$_2$H$_2$ ice at 15~K, taken as 1.34 \citep{Hudson_Ferrante_Moore_2014, Abplanalp_Kaiser_2020}, and \(\theta_{\text{ice}}\) is the angle of incidence (2$^\circ$). The ice density values of 0.76~g~cm$^{-3}$ \citep{Hudson_Ferrante_Moore_2014}, reported in the literature are used to convert the measured absolute ice thicknesses into the corresponding column densities. The molecule deposition rate is derived by dividing the obtained column density by its deposition time.

 All experiments were performed using high-purity (4N) C$_2$H$_2$ (Linde 5\% in helium). The grown C$_2$H$_2$ ice samples are exposed to VUV photons produced by a microwave discharge hydrogen lamp (MDHL) and guided through a magnesium fluoride (MgF$_2$) viewport. The estimated flux of the applied MDHL is (3 ± 0.5)$\times$10$^{14}$~photons~cm$^{-2}$~s$^{-1}$ at the substrate plane characterized with two main emission bands at 121.6~nm (10.2~eV) and 140--170~nm (7.2-8.9~eV) \citep{Ligterink_Paardekooper_Chuang_Both_Cruz_Diaz_Van_Helden_Linnartz_2015}, mimicking the secondary UV emissions produced by the interaction of cosmic rays with the abundant H$_2$ in molecular clouds \citep{Prasad_Tarafdar_1983}. 

\begin{figure}[h]
   \centering
     \includegraphics[width = 0.7\linewidth]{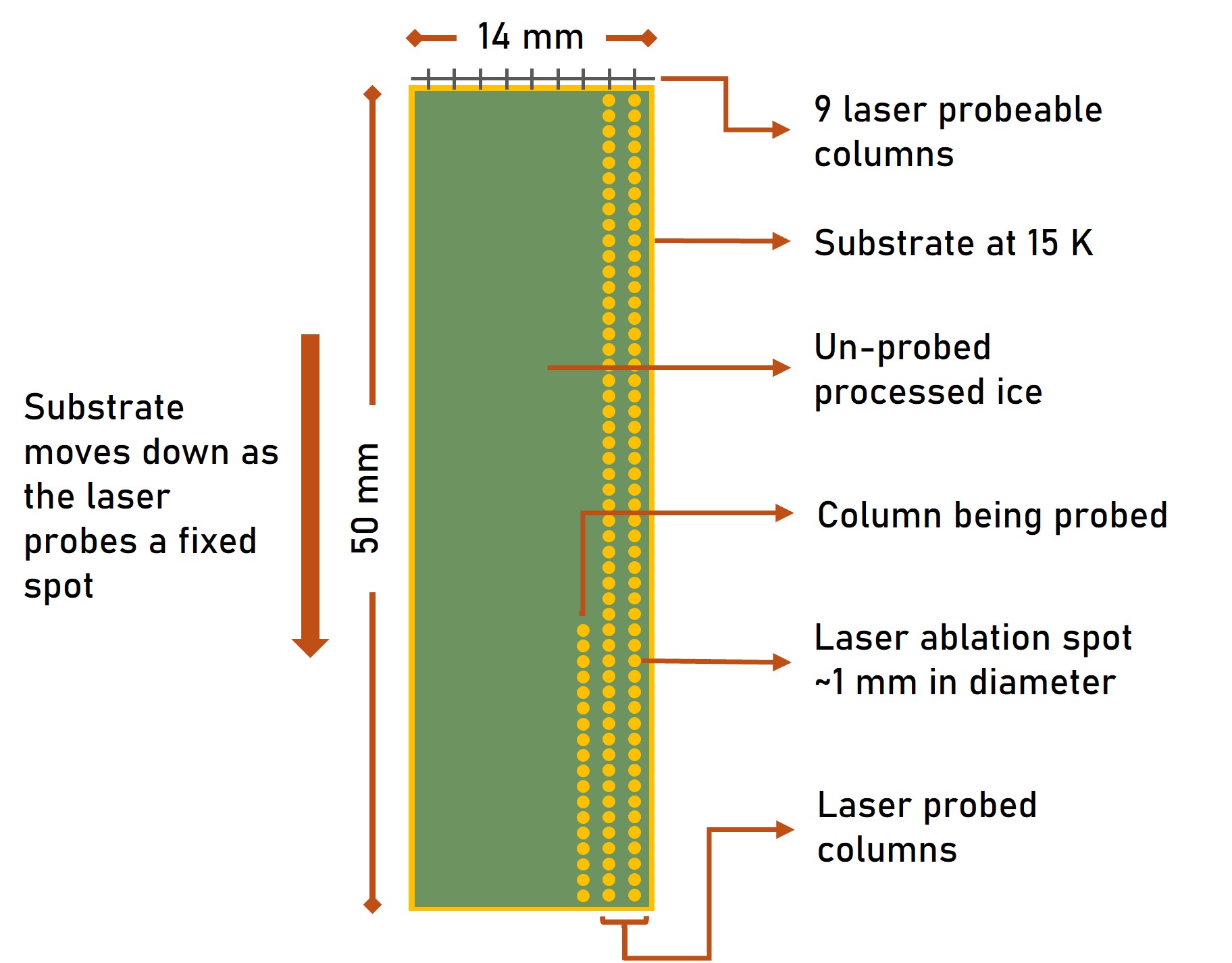}
    \caption{Schematic illustration of the ablation scheme used at the MATRI$^2$CES setup. The laser fires at 5 Hz while the substrate is moved downward by a motorized Y-translator, ensuring a fresh spot is probed with each shot.}
     \label{fig:fig001}
\end{figure}

\subsection{\label{sec:level22}LDPI ReTOF-MS technique}

After VUV processing, a portion of the ice is probed by pulsing a Nd:YAG laser (New Wave Research, Polaris II) at its third harmonic (355~nm) onto the substrate. The laser pulse locally desorbs a small spot (approximately 1 mm in diameter, as set by a manually controlled aperture) from the substrate, transferring the ice constituents into the gas phase via flash heating in a short pulse of 3–5~ns. This ablation produces a short-lived ($\mu$s) gas plume \citep{Henderson_Gudipati_2014}, which is immediately ionized by a continuously operating, tunable electron beam (Jordan C-950), positioned in parallel to the substrate. This electron energy is set to 70~eV to obtain corresponding mass fragments that can be directly compared with the National Institute of Standards and Technology (NIST) database \citep{NIST_ChemWebBook, Bulak_Paardekooper_Fedoseev_Linnartz_2021}. The generated ions are then guided into the ToF chamber, operated in the reflection mode to enhance mass resolution, by a set of ion optics, including a repeller plate, extraction grid, and deflection plate which work together to direct the ions into the field-free drift tube and travel toward a micro-channel plate (MCP) detector. The detected ion signals are recorded using a data acquisition card (DAQ) at a sampling rate of 2.5~GHz. The laser pulse, ion extraction, and data acquisition sequence are controlled through a delay generator (Stanford DG-535) and the acquired data is processed through a LabVIEW program. During laser ablation at 5~Hz, the substrate is moved vertically, in sync with a stepper motor-controlled linear translator, so that each laser shot probes a fresh spot on the substrate. In each column measurement, around 1 mm wide, 100 mass spectra are collected and averaged into a single mass spectrum to improve the signal-to-noise ratio. Once a measurement is complete, the substrate returns to its original vertical position and is moved horizontally by 1.5~mm using a hand-cranked translator so that the laser probing process can be repeated for another fresh column of ice. This experimental procedure enables the acquisition of mass spectra for multiple columns across the substrate as a function of VUV irradiation fluence without repeating the experiment for different UV irradiation times. Given that the substrate's width is 14 mm, in total, 9 mass spectra (columns) can be probed and collected during the irradiation of the same ice sample. The substrate, probed columns, and the laser-probed spots are illustrated in Fig.~1.

Additionally, with a recent upgrade to MATRI$^2$CES, a pulsed ion deflection (PID) method was implemented to overcome the signal saturation limitation of the ReTOF-MS. Signal saturation occurs when the high abundance of the parent molecular ions causes the detector to saturate at an intensity of approximately 1~V. This hinders the detection of the newly formed species produced in much smaller quantities. With the PID method, this saturation is overcome by applying a variable short voltage (50~V) pulse to the deflection plates, controlled by a second delay generator (Stanford DG 535). By precisely controlling the deflection time between 0.2--1~$\mu$s, the most abundant m/z signals are selectively deflected away from reaching the MCP detector. This technique has been experimentally shown to increase the MCP sensitivity by approximately 30 times in a previous study on the UV photolysis of methanol (CH$_3$OH) ice sample \citep{Samarth_Bulak_Paardekooper_Chuang_Linnartz_2024}.

\section{\label{sec:level3} Results and discussion \protect}

\subsection{\label{sec:level31} LDPI ReTOF-MS of pure C$_2$H$_2$ ice }

\begin{figure*}[t]
   \centering
     \includegraphics[width = 1.2\linewidth]{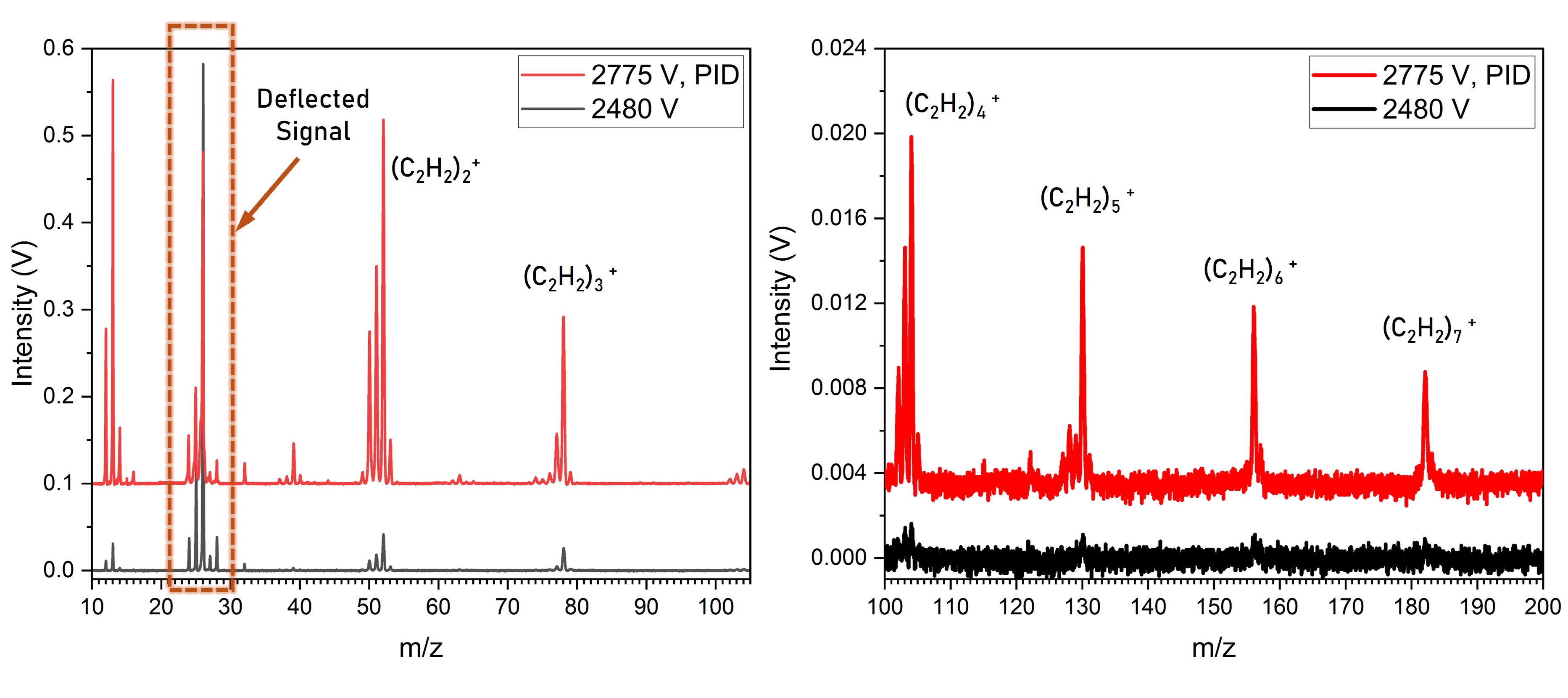}
    \caption{LDPI ReTOF-mass spectra recorded with PID at an increased voltage of 2775 V (Red) and without PID at a base voltage of 2480 V (Black). Both spectra are obtained for 100~ML thick unirradiated C$_2$H$_2$ ice at 15~K. The high sensitivity mass spectrum shows a wealth of spectral features belonging to C$_2$H$_2$ clusters at masses m/z = 52 (C$_2$H$_2$)$_2$$^+$, 78 (C$_2$H$_2$)$_3$$^+$, 104 (C$_2$H$_2$)$_4$$^+$, 130 (C$_2$H$_2$)$_5$$^+$, 156 (C$_2$H$_2$)$_6$$^+$, and 182 (C$_2$H$_2$)$_7$$^+$.}
     \label{fig:fig002}
\end{figure*}

The averaged LDPI ReTOF mass spectrum of pure C$_2$H$_2$ ice 100~ML thick at 15~K, obtained with and without PID setting, is shown in Fig.~2. The reference mass spectrum (Fig.~2, black trace) collected at a sensitivity of 2480~V, shows the fragmentation pattern of pure C$_2$H$_2$ ice using 70~eV ionization energy displays peaks at m/z = 12 (C$^+$), 13 (CH$^+$), 14 (CH$_2$$^+$), 24 (C$_2$$^+$), 25 (C$_2$H$^+$), and 26 (C$_2$H$_2$$^+$). In addition, the ion signals of C$_2$H$_2$ clusters (C$_2$H$_2$)$_n$$^+$, at masses m/z = 52 and m/z = 78 for n = 2 and 3 are evident. Hereafter, we refer to the detector voltage readings as the intensity.

In the present work, PID was implemented by applying a precisely controlled voltage pulse of 0.5 $\mu$s with a calibrated delay of 2.426 $\mu$s after the ion extraction was triggered. The trigger timing and duration are set to deflect the majority of C$_2$H$_2$ fragment signals between masses m/z = 23 and 27 without affecting the rest of the mass spectrum. As a result, the mass spectrum exhibits significantly reduced intensities for these parent ions, enabling an increase in the MCP bias voltage from 2480 V to 2775 V. The resulting mass spectrum (Fig.~2, red trace) shows a notable decrease in the intensity of parent ion peaks between m/z = 23 and 27 relative to their intensities in the reference spectrum, even at a higher sensitivity of 2775 V. Moreover, the PID enhanced spectrum reveals not only an increase in the intensity of previously observed peaks belonging to C$_2$H$_2$ clusters at masses m/z = 52 and 78, but also the appearance of a higher-order (C$_2$H$_2$)$_n$$^+$ cluster peaks at m/z = 104, 130, 156, and 182 for n = 4, 5, 6, and 7, respectively. The assignment of these C$_2$H$_2$ clusters is confirmed by performing a series of controlled experiments, which are summarized in Fig.~A.1 in Appendix~A. Similar C$_2$H$_2$ cluster formations have been observed in previous studies of post-ionized supersonic C$_2$H$_2$ beam expansion with small signal at masses m/z = 37, 38, 39, and 63 \citep{Momoh_Abrash_Mabrouki_Samy_El-Shall_2006, El-Shall_2008, Relph_Bopp_Roscioli_Johnson_2009}.

\begin{figure*}[htp]
   \centering
     \includegraphics[width = 1.2\linewidth]{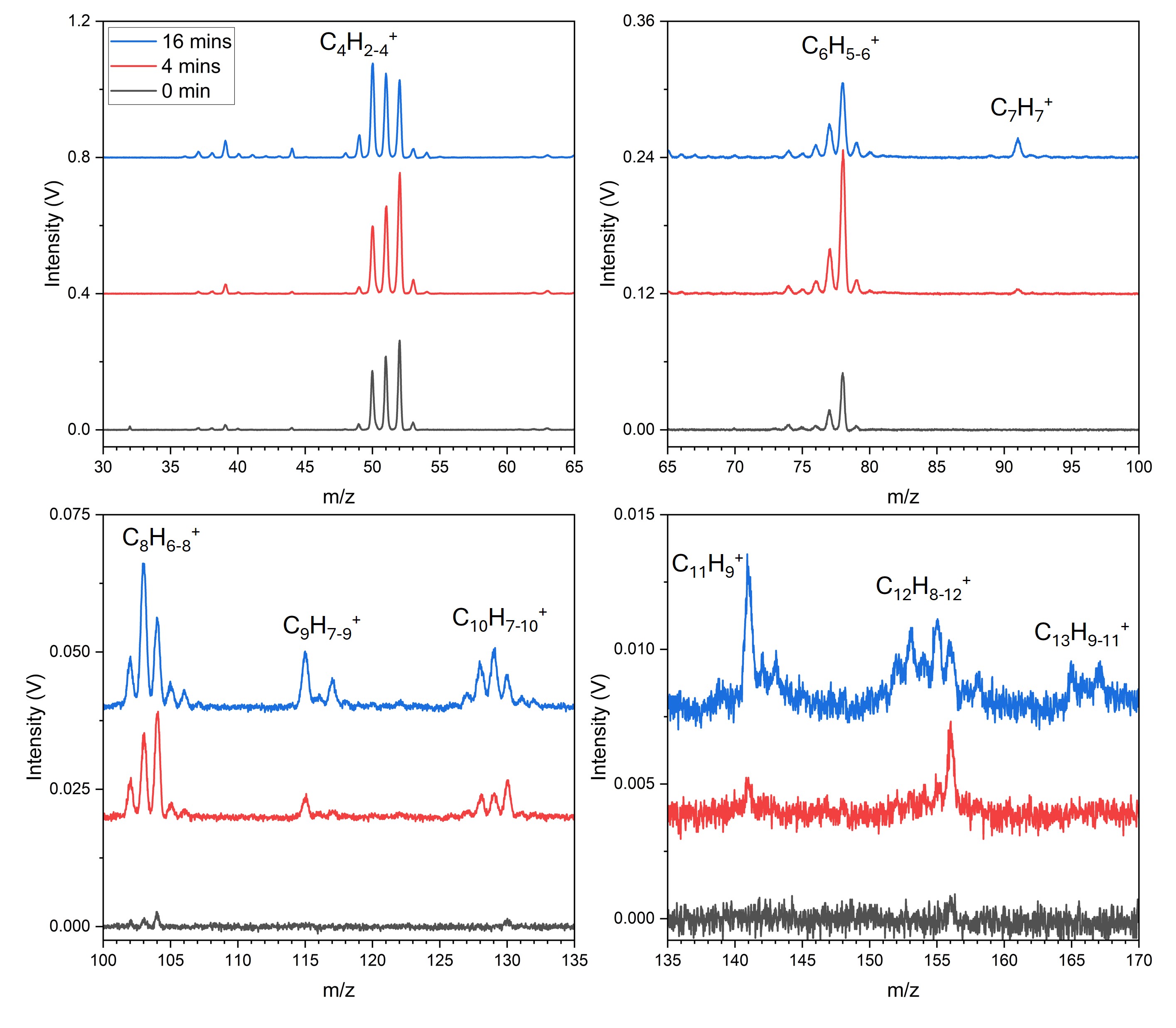}
           \caption{LDPI ReTOF mass spectra obtained after VUV irradiation of 100~ML thick C$_2$H$_2$ ice with varying VUV photon exposures of $7.2 \times 10^{16}$~photons~cm$^{-2}$ (4 minutes) and $3 \times 10^{17}$~photons~cm$^{-2}$ (16 minutes) are compared to the  unirradiated mass spectra at 15~K. The spectra are offset for clarity. The complete list of detected mass signals is provided in Table~1.
}
     \label{fig:fig003}
\end{figure*}

\subsection{\label{sec:level32} VUV irradiation of pure C$_2$H$_2$ ice at 15~K}

The C$_2$H$_2$ ice samples were irradiated at 15~K for up to 64 minutes of VUV irradiation. This corresponds to a VUV-photon fluence of $\sim$1.2 $\times$ 10$^{18}$ photons cm$^{-2}$. During irradiation, changes in the ice were monitored by recording mass spectra at fixed time intervals of 4, 8, 16, 32, and 64 minutes. Fig.~3 shows the selected mass spectra obtained after 4 (red) and 16 (blue) minutes of irradiation of 100 ML-thick pure C$_2$H$_2$ ice, together with the mass spectrum of  unirradiated C$_2$H$_2$ ice as a reference (black). These irradiation times correspond to an exposure of $\sim$7 $\times$ 10$^{16}$ photons cm$^{-2}$ and $\sim$3 $\times$ 10$^{17}$ photons cm$^{-2}$, respectively. The UV photolysis of pure C$_2$H$_2$ ice can lead to both photodesorption and photoconversion at 20~K. The 16-minute irradiation spectrum shows the maximum intensity of photoproduct signals within the recorded mass range between m/z = 30 and 170, while the 4-minute irradiation spectrum provides a representative snapshot of the early photochemistry of primary and secondary ice constituents.

In the 4-minute (red trace) irradiation spectrum in Fig.~3, a rapid formation of photoproducts with an even number of C atoms can be observed in the mass ranges between m/z = 50 and 52 (C$_4$H$_{2-4}$), 77 and 78 (C$_6$H$_{5-8}$), 102 and 104 (C$_8$H$_{6-8}$), 127 and 130 (C$_{10}$H$_{7-10}$), and 152 and 156 (C$_{12}$H$_{8-12}$).
 This observation is in line with the efficient radical polymerization (or cyclization) reactions, reported by \citet{Cuylle_Zhao_Strazzulla_Linnartz_2014}. In addition to hydrocarbons with even numbers of carbon atoms, relatively weak signals corresponding to odd-carbon hydrocarbons are also observed at m/z = 91 (C$_7$H$_{7}$), m/z = 115 (C$_9$H$_{7}$), m/z = 141 (C$_{11}$H$_{9}$), and m/z = 165 (C$_{13}$H$_{9}$) suggesting a slower formation process. As these photoproducts consist exclusively of elemental carbon and hydrogen originating from C$_2$H$_2$, the most intense peaks can be assigned according to their C/H compositions and are labeled in Fig.~3. A complete assignment of all peaks is provided in Table~1. As C$_2$H$_2$ ice is progressively depleted upon VUV irradiation, the overall abundance of (C$_2$H$_2$)$_n^+$ clusters is expected to decline with it. In our mass spectra, where the intensities of photoproduct mass fragments (Fig.~3, red and blue traces) significantly exceed those of the unirradiated C$_2$H$_2$ clusters (Fig.~3, black trace), we do not consider them in our analysis.

Table~1 lists all photoproducts detected after 16 minutes of irradiation of pure C$_2$H$_2$ at 15~K, sorted by carbon atom count (first column), m/z values in descending order (second column), and corresponding elemental compositions (third column). The relatively intense peaks (intensity >50\% compared to the strongest peak in each group) are further highlighted in boldface. The fourth column shows the aromatic molecules unambiguously identified by \citet{Abplanalp_Kaiser_2020} through application of REMPI ReTOF-MS and the fifth column shows other molecules suggested in this work based on the identifications of \citet{Abplanalp_Kaiser_2020}. This study confirms the presence of all parent molecular mass signals corresponding to aromatic hydrocarbons, including m/z = 78 (C$_6$H$_6$, benzene), m/z = 102 (C$_8$H$_6$, phenylacetylene), m/z = 104 (C$_8$H$_8$, phenylethylene), and m/z = 128 (C$_{10}$H$_8$, naphthalene), as previously shown by REMPI ReTOF-MS after 5~keV electron irradiation of C$_2$H$_2$ ice at 5~K.

Besides these confirmed hydrocarbons, the intense signals at m/z = 50, 51, 52 can be securely assigned to C$_4$H$_2$ (diacetylene) and C$_4$H$_4$ (vinylacetylene) molecules, which are two common chemical derivatives originating from C$_2$H$_2$$^+$ \citep{Willis_Back_Morris_1977, Pereira_deBarros_daCosta_Oliveira_Fulvio_daSilveira_2020, Chuang_Fedoseev_Scirè_Baratta_Jäger_Henning_Linnartz_Palumbo_2021}. Similarly, the assignment of m/z = 78, 77, 79 signals to linear hexadienyne (C$_6$H$_6$), along with the previous assignment to cyclic benzene, can be considered. It is noted that LDPI ReTOF-MS alone does not allow for the discrimination between the linear and cyclic conjugated unsaturated hydrocarbons due to the possible internal isomerization of ions analyzed by TOF-MS. Therefore, hereafter, the preference is given to the assignments of molecules identified by \cite{Abplanalp_Kaiser_2020} using REMPI ReTOF-MS, and to the search for their possible derivatives.

\begin{table*}[htp]
\centering
\renewcommand{\arraystretch}{1.5}
\setlength{\tabcolsep}{4pt}
\caption{List of detected photoproducts.}
\label{tab:mass_peaks_pah_analysis}
\begin{tabular}{>{\raggedright\arraybackslash}p{0.7cm} >{\raggedright\arraybackslash}p{1.6cm} >{\raggedright\arraybackslash}p{2.6cm} >{\raggedright\arraybackslash}p{4.3cm} >{\raggedright\arraybackslash}p{7.5cm}}

\hline\hline
No. of C & m/z (decreasing intensity) & Corresponding elemental composition & Aromatic molecules identified using REMPI ReTOF-MS \citep{Abplanalp_Kaiser_2020} & Other molecules proposed based on the identifications of \citet{Abplanalp_Kaiser_2020} and analysis of our LDPI ReTOF-MS \\
\hline
3  & \textbf{39}\footnotemark[1] & \textbf{C$_3$H$_3$$^+$} &  & m/z = 40, 39 - C$_3$H$_4$ (propyne), R–C$_3$H$_3$ (propynyl group) \\
4  & \textbf{50}, \textbf{51}, \textbf{52}, 49, 53, 54 & \textbf{C$_4$H$_2$$^+$}, \textbf{C$_4$H$_3$$^+$}, \textbf{C$_4$H$_4$$^+$}, C$_4$H$^+$, C$_4$H$_5$$^+$, C$_4$H$_6$$^+$ &  & m/z = 50, 49 - C$_4$H$_2$ (diacetylene); m/z = 52, 51, 50 - C$_4$H$_4$ (vinylacetylene) \\
6  & \textbf{78}, \textbf{77}, 79, 76 & \textbf{C$_6$H$_6$$^+$}, \textbf{C$_6$H$_5$$^+$}, C$_6$H$_7$$^+$, C$_6$H$_4$$^+$ & m/z = 78, 77 - C$_6$H$_6$ (benzene) & m/z = 78, 77 - C$_6$H$_6$ (benzene), m/z = 78 - C$_6$H$_6$ (linear - hexadienyne) \\
7  & \textbf{91} & \textbf{C$_7$H$_7$$^+$} &  & m/z = 91, 92 - C$_6$H$_5$CH$_3$ (methylbenzene), C$_6$H$_5$CH$_2$–R (substituted methylbenzene) \\
8  & \textbf{103}, \textbf{104}, \textbf{102}, 105, 106 & \textbf{C$_8$H$_7$$^+$}, \textbf{C$_8$H$_8$$^+$}, \textbf{C$_8$H$_6$$^+$}, C$_8$H$_9$$^+$, C$_8$H$_{10}$$^+$ & m/z = 104, 103 - C$_6$H$_5$C$_2$H$_3$ (phenylethylene); m/z = 102  - C$_6$H$_5$CCH (phenylacetylene) &  \\
9  & \textbf{115}, 117 & \textbf{C$_9$H$_7$$^+$}, C$_9$H$_9$$^+$ &  & m/z = 115 - R–CH$_2$C$_6$H$_4$CCH or C$_6$H$_5$CCCH$_2$–R; m/z = 117 - R–CH$_2$C$_6$H$_4$C$_2$H$_3$ or C$_6$H$_5$C$_2$H$_2$–R (substituted phenylethylenes and phenylacetylenes) \\
10 & \textbf{129}, \textbf{128}, \textbf{130}, 127 & \textbf{C$_{10}$H$_9$$^+$}, \textbf{C$_{10}$H$_8$$^+$}, \textbf{C$_{10}$H$_{10}$$^+$}, C$_{10}$H$_7$$^+$ & m/z = 128, 129, 127 - C$_{10}$H$_8$ (naphthalene) &  \\
11 & \textbf{141} & \textbf{C$_{11}$H$_9$$^+$} &  & m/z = 141 - C$_{10}$H$_7$CH$_2$–R or R–C$_{10}$H$_6$CH$_3$ (substituted naphthalenes) \\
12 & \textbf{155}, \textbf{153}, \textbf{156}, 152, 154 & \textbf{C$_{12}$H$_{11}$$^+$}, \textbf{C$_{12}$H$_9$$^+$}, \textbf{C$_{12}$H$_{12}$$^+$}, C$_{12}$H$_8$$^+$, C$_{12}$H$_{10}$$^+$ &  & m/z = 155 - C$_{10}$H$_7$C$_2$H$_4$–R or C$_{10}$H$_7$CH(-R)CH$_2$; m/z = 153 - C$_{10}$H$_7$C$_2$H$_2$–R \\
13 & \textbf{165}, \textbf{167} & \textbf{C$_{13}$H$_9$$^+$}, \textbf{C$_{13}$H$_{11}$$^+$} &  & m/z = 165 - C$_{13}$H$_9$–R (phenalene derivatives) \\
\hline\hline
\end{tabular}
\tablefoot{List of photoproducts after VUV irradiation of pure C$_2$H$_2$ ice at 15~K for a total fluence of 3 $\times$ 10$^{17}$ photons cm$^{-2}$ (16 minutes). The signals are grouped according to the total number of C atoms.}
\end{table*}
\footnotetext[1]{The m/z values with the relative intensity of >50\% are in bold font.}

The intense signals at m/z = 91, 115, 117, 141, 153, 155, and 165 are abundantly observed in the post-16-minXute irradiation spectrum. Based on the structure of PAHs identified in \cite{Abplanalp_Kaiser_2020}, several other substituted PAHs with C$_n$H$_m$ side groups, such as CH$_3$, –C$_2$H, –C$_2$H$_3$, –C$_2$H$_5$, –C$_3$H$_3$, can be expected, for example, methylbenzene (toluene, C$_6$H$_5$CH$_3$), phenylmethylacetylene (C$_6$H$_5$CCCH$_3$), methylnaphthalene (C$_{10}$H$_7$CH$_3$), vinylnaphthalene (C$_{10}$H$_7$C$_2$H$_3$), ethylnaphthalene (C$_{10}$H$_7$C$_2$H$_5$) with molecular ions at m/z = 92, 116, 142, 154, 156, respectively. It is noted that a systematic m/z = 1 shift exists between the aforementioned species and the mass signals observed in the experiments. This is consistent with a loss of a single H atom upon dissociative ionization or the loss of any other side-substituted (-R) groups, hinting at the presence of complicated aliphatic chains attached to these observed PAH rings. For example, the mass signal at m/z = 91, the tropylium ion (C$_7$H$_7$), is a major ion fragment in the mass spectra of many substituted arenes (hydrocarbons that contain at least one benzene ring), including ethylbenzene (C$_8$H$_{10}$), butylbenzene (C$_{10}$H$_{14}$), and xylene (C$_8$H$_{10}$).

The most intense signals observed in the mass range of m/z = 76 and 80 suggest the presence of aromatic structures based on the mass spectra available on the NIST database. This can be illustrated by comparing the mass fragmentation patterns of a PAH mentioned in this work, naphthalene (C$_{10}$H$_8$, m/z = 128), which exhibits a dominant peak at m/z = 128 with only minor dissociative ionization fragments in the lower mass range, with its isobaric aliphatic counterpart, nonane (C$_9$H$_{20}$, m/z = 128) which yields a cascade of lower-mass fragments, most of which are not seen in our collected mass spectrum with negligible signal at m/z = 128; shown in Appendix~B. The intense mass signal at m/z = 115 (C$_9$H$_7$) corresponds to the most intense mass fragment signal of substituted indenes, a benzene ring fused with a cyclopentene ring. Similarly, the most intense mass signal at m/z = 165 (C$_{13}$H$_9$) corresponds to the phenalene derivative (C$_{13}$H$_{10}$), one of the simplest true PAHs. It is important to note that, because LDPI ReTOF–MS is not optimized to record mass spectra of the ice sample during irradiation and cannot resolve the overlapping thermal desorption temperatures of the complex suite of photoproducts, these assignments are considered tentative. A quantitative assessment of photodesorption and photoconversion, as well as a detailed TPD analysis, is beyond the scope of the present work and will require a dedicated future study.

\subsection{\label{sec:level33}Refractory residues at 300~K}

\begin{figure*}[htp]
   \centering
     \includegraphics[width = 1.1\linewidth]{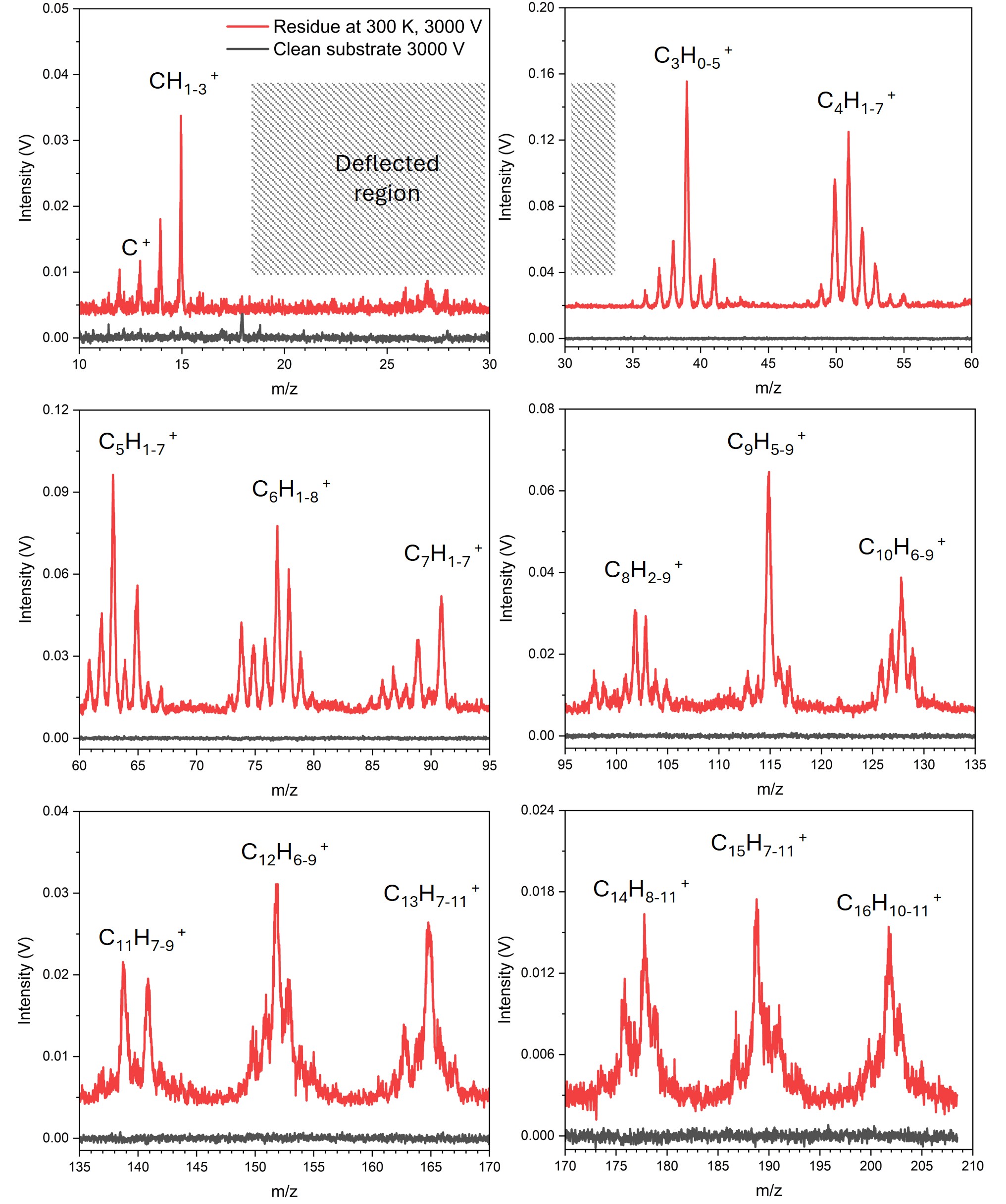}
    \caption{LDPI ReTOF-MS of the organic residue produced by VUV irradiation of 100 ML thick C$_2$H$_2$ with a total fluence of 3 $\times$ 10$^{17}$ photons cm$^{-2}$ fluence at 15~K. The mass spectrum is acquired at 300~K upon the annealing of the ice at an increased detector voltage of 3000 V. The complete list of detected mass signals is provided in Table~2. The deflected region is the mass range between m/z = 18 and 34 where the ions are deflected upon the use of the PID technique.}
     \label{fig:fig004}
\end{figure*}

Two types of dedicated experiments were performed to probe the nonvolatile organic residues produced by the irradiation of C$_2$H$_2$ ice at 15~K. In the first type of experiment, the irradiated C$_2$H$_2$ ice samples obtained at 15~K are gently warmed to 300~K at a constant rate of 2.5~K min$^{-1}$ and kept isotherm until the main chamber pressure reaches the base 10$^{-10}$ mbar range. Consequently, the LDPI ReTOF mass spectrum of the refractory residue was acquired as shown in Fig.~4. In the second type of experiments, the obtained refractory residue is kept overnight at 300~K in order to achieve the complete degassing of the obtained residue and elimination of the desorbed species from the gas phase. Then the refractory residues were cooled down to 15~K, and the LDPI ReTOF mass spectra were obtained. In this way, the mass spectra with the lowest possible contribution of the main chamber background gas constituents are obtained; shown in Appendix~D Fig.~D,1. In both sets of experiments, the MCP detector bias voltage was increased from 2775 V to 3000 V to enhance the mass signals of the present residue, as due to the absence of parent ice and volatile product constituents after annealing, there was no risk of detector saturation. It was found that both measurements result in the acquisition of qualitatively very similar mass spectra. The mass spectra collected after waiting overnight carry an additional mass signal of H$_2$O. Moreover, the obtained mass spectra of the nonvolatile residues are qualitatively similar for all of the applied experimental conditions.

\begin{table*}[ht]
\centering
\renewcommand{\arraystretch}{1.5}
\setlength{\tabcolsep}{8pt} % Added to increase spacing between columns
\renewcommand{\arraystretch}{1.15} % Increased row spacing
\caption{List of detected organic residues.}
\label{tab:stoichiometry_assignments_2}

\begin{tabular}{>{\raggedright\arraybackslash}p{0.7cm} >{\raggedright\arraybackslash}p{2.3cm}
>{\raggedright\arraybackslash}p{4cm}
>{\raggedright\arraybackslash}p{7.5cm}}
\hline\hline
No. of C & m/z (decreasing intensity) & Corresponding elemental composition & Possible key substituted functional groups proposed in this work \\
\hline
1 & \textbf{15}, 14, 13, 12 & \textbf{CH$_3^+$}, CH$_2^+$, CH$^+$, C$^+$ & m/z = 15 - R--CH$_3$ \\
3 & \textbf{39}, 38, 41, 40, 37, 36 & \textbf{C$_3$H$_3$$^+$}, C$_3$H$_2$$^+$, C$_3$H$_5$$^+$, C$_3$H$_4$$^+$, C$_3$H$_1$$^+$, C$_3$$^+$ & m/z = 39 = R--CH$_2$--C$\equiv$CH or R--C$\equiv$C--CH$_3$ or R--CH=C=CH$_2$ \\
4 & \textbf{51}, \textbf{50}, \textbf{52}, 53, 49, 54, 55 & \textbf{C$_4$H$_3$$^+$}, \textbf{C$_4$H$_2$$^+$}, \textbf{C$_4$H$_4$$^+$}, C$_4$H$_5^+$, C$_4$H$_1$$^+$, C$_4$H$_6$$^+$, C$_4$H$_7$$^+$ & m/z = 51 = R--CH=CH--C$\equiv$CH or R--C$\equiv$C--CH=CH$_2$, m/z = 53 = R--CH=CH--CH=CH$_2$ \\
5 & \textbf{63}, \textbf{65}, 62, 61, 64, 66, 67 & \textbf{C$_5$H$_3$$^+$}, \textbf{C$_5$H$_5$$^+$}, C$_5$H$_2$$^+$, C$_5$H$_1$$^+$, C$_5$H$_4$$^+$, C$_5$H$_6$$^+$, C$_5$H$_7$$^+$ & m/z = 63 = R--CH$_2$--C$\equiv$C--C$\equiv$CH or R--C$\equiv$C--CH$_2$--C$\equiv$CH or R--C$\equiv$C--C$\equiv$C--CH$_3$, m/z = 65 = R--C$\equiv$C--CH=CH--CH$_3$ or R--CH=CH--CH$_2$--C$\equiv$CH or R--CH$_2$--CH=CH--C$\equiv$CH, etc. \\
6 & \textbf{77}, \textbf{78}, 74, 76, 75, 79, 80 & \textbf{C$_6$H$_5$$^+$}, \textbf{C$_6$H$_6$$^+$}, C$_6$H$_2$$^+$, C$_6$H$_4$$^+$, C$_6$H$_3$$^+$, C$_6$H$_7$$^+$, C$_6$H$_8$$^+$ & m/z = 77 = R--C$_6$H$_5$ (phenyl group) or R--CH=CH--CH=CH--C$\equiv$CH or R--C$\equiv$C--CH=CH--CH=CH$_2$ or m/z = 78 (Benzene), etc.\\
7 & \textbf{91}, \textbf{89}, 87, 86, 88 & \textbf{C$_7$H$_7$$^+$}, \textbf{C$_7$H$_5$$^+$}, C$_7$H$_3$$^+$, C$_7$H$_2$$^+$, C$_7$H$_4$$^+$ & m/z = 91 = R--CH$_2$--C$_6$H$_5$ (benzyl group), m/z = 89 = R--CH$_2$--CH$\equiv$CH--CH=CH--C$\equiv$CH, etc. \\
8 & \textbf{102}, \textbf{103}, 98, 104, 105, 101, 99 & \textbf{C$_8$H$_6$$^+$}, \textbf{C$_8$H$_7$$^+$}, C$_8$H$_2$$^+$, C$_8$H$_8$$^+$, C$_8$H$_9$$^+$, C$_8$H$_5$$^+$, C$_8$H$_3$$^+$ & m/z = 102 = H--C$\equiv$C--C$_6$H$_5$ (phenylacetylene), m/z = 103 = R--CH=CH--C$_6$H$_5$ (substituted styrenes) \\
9 & \textbf{115}, 116, 117, 113, 114 & \textbf{C$_9$H$_7$$^+$}, C$_9$H$_8$$^+$, C$_9$H$_9$$^+$, C$_9$H$_5$$^+$, C$_9$H$_6$$^+$ & m/z = 115 = R--CH$_2$--C$\equiv$C--C$_6$H$_5$ or R--C$_9$H$_7$ (substituted indenes) \\
10 & \textbf{128}, \textbf{127}, 129, 126 & \textbf{C$_{10}$H$_8$$^+$}, \textbf{C$_{10}$H$_7$$^+$}, C$_{10}$H$_9$$^+$, C$_{10}$H$_6$$^+$ & m/z = 128 = C$_{10}$H$_8$ (naphthalene), 127 = R--C$_{10}$H$_7$ or R--CH=CH--C$\equiv$C--C$_6$H$_5$ \\
11 & \textbf{139}, \textbf{141} & \textbf{C$_{11}$H$_7$$^+$}, \textbf{C$_{11}$H$_9$$^+$} & m/z = 139 = R--CH$_2$--C$\equiv$C--C$\equiv$C--C$_6$H$_5$, m/z = 141 = R--CH$_2$--C$_{10}$H$_7$ \\
12 & \textbf{152}, 153, 151, 150, 154 & \textbf{C$_{12}$H$_8$$^+$}, C$_{12}$H$_9$$^+$, C$_{12}$H$_7$$^+$, C$_{12}$H$_6$$^+$, C$_{12}$H$_{10}$$^+$ & m/z = 152 = HC$\equiv$C--C$_{10}$H$_7$ or C$_{12}$H$_8$ (acenaphthylene?) \\
13 & \textbf{165}, 163 & \textbf{C$_{13}$H$_9$$^+$}, C$_{13}$H$_7$$^+$ & m/z = 165 = R--CH$_2$--C$\equiv$C--C$_{10}$H$_7$ or R--C$_{13}$H$_9$ (substituted phenalenes?) \\
14 & \textbf{178}, 176 & \textbf{C$_{14}$H$_{10}$$^+$}, C$_{14}$H$_8$$^+$ & m/z = 178 = C$_{14}$H$_{10}$ (phenanthrene) or H$_2$C=CH--C$\equiv$C--C$_{10}$H$_7$ \\
15 & \textbf{189}, 191, 187 & \textbf{C$_{15}$H$_9$$^+$}, C$_{15}$H$_{11}$$^+$, C$_{15}$H$_7$$^+$ & m/z = 189 = R--CH$_2$--C$\equiv$C--C$\equiv$C--C$_{10}$H$_7$ \\
16 & \textbf{202}, 203, 201 & \textbf{C$_{16}$H$_{10}$$^+$}, C$_{16}$H$_{11}$$^+$ & m/z = 202 = C$_{16}$H$_{10}$ (pyrene), HC$\equiv$C--C$_{14}$H$_9$ \\
\hline\hline
\end{tabular}
\tablefoot{List of detected organic residues after VUV irradiation of pure C$_2$H$_2$ ice at 15~K for a total fluence of 3 $\times$ 10$^{17}$ photons cm$^{-2}$ (16 minutes). A list of ion signals observed in the LDPI ReTOF mass spectra of the organic residue presented in Fig.~4.}
\end{table*}

Fig.~4 shows the mass spectra of the refractory organic residue at 300~K obtained after irradiation of 100~ML thick pure C$_2$H$_2$ ice with a total fluence of 3 $\times$ 10$^{17}$~photons~cm$^{-2}$. The main refractory peaks are grouped and labeled with their corresponding chemical formulas. A full list of peak assignments is provided in Table~2. Similar to Table~1, ion signals are organized by the number of carbon atoms (first column), followed by m/z values listed in descending order of intensity (second column). The third column shows the corresponding chemical formulas in the same order. The fourth column suggests possible key functional groups or fragments, assuming they result from the dissociative ionization of larger molecules present in the residue.

The residue mass spectrum exhibits a series of mass peak groups with an average separation of 12 to 13 m/z units, which corresponds to the mass of a C atom or a -CH- group. Such periodicity suggests that the produced residue predominantly consists of conjugated unsaturated hydrocarbon units, characterized by alternating triple bonds (-C$\equiv$C-) and double bonds (-CH=CH-) with segments of m/z = 24 and m/z = 26, respectively. In Fig.~4, mass peaks between m/z = 30 and 105 display a broader variation in hydrogenation levels by up to 8 m/z units. This hints at the presence of several aliphatic unsaturated fragments alongside aromatic fragments in the mass spectrum. For example, the presence of peaks at m/z = 50 and m/z = 74 can be attributed to the polyynic structures such as C$_4$H$_2$ and C$_6$H$_2$, respectively. While the peaks at m/z = 51 and m/z = 75 can be assigned to polyenyne fragments, such as C$_4$H$_3$ and C$_6$H$_3$, for example, aliphatic fragments comprised of both triple and double carbon-carbon bonds. Among the odd carbon number ions, the most intense signals correspond to C$_3$H$_3$ (m/z = 39), C$_5$H$_3$ (m/z = 63), and C$_7$H$_3$ (m/z = 87). Appearance of these signals is in line with the presence of methylpolyyne segments in the parent ionized species. Additionally, the characteristic m/z signals corresponding to the resonance‐stabilized cyclic ions are evident, with notable peaks including aforementioned C$_3$H$_3$ at m/z = 39, C$_6$H$_5$ at m/z = 65, and C$_7$H$_7$ at m/z = 91. Although the former ion may also originate from the aliphatic propargyl (-CH$_2$-C$\equiv$CH), propynyl (-C$\equiv$C-CH$_3$) or even allenyl (-HC=C=CH$_2$) groups. 

Beyond m/z = 105, the degree of hydrogenation decreases, and the observed peak groups tend to be dominated by a few, well-defined peaks. This observation hints at the presence of polycyclic aromatic hydrocarbons (PAHs) and their substituted derivatives. For instance, the strong signal at m/z = 128 corresponds to naphthalene (C$_{10}$H$_8$), a well-known aromatic hydrocarbon. Similarly, the m/z = 178 and 202 signals, the most intense within their groups, correspond to phenanthrene (or anthracene) and pyrene, respectively. The m/z signals of 115 and 165 can be assigned to indenyl (C$_{9}$H$_7$, m/z = 115) and phenalenyl (C$_{13}$H$_9$, m/z = 165), respectively.

\begin{figure*}[htp]
   \centering
     \includegraphics[width = 1.2 \linewidth]{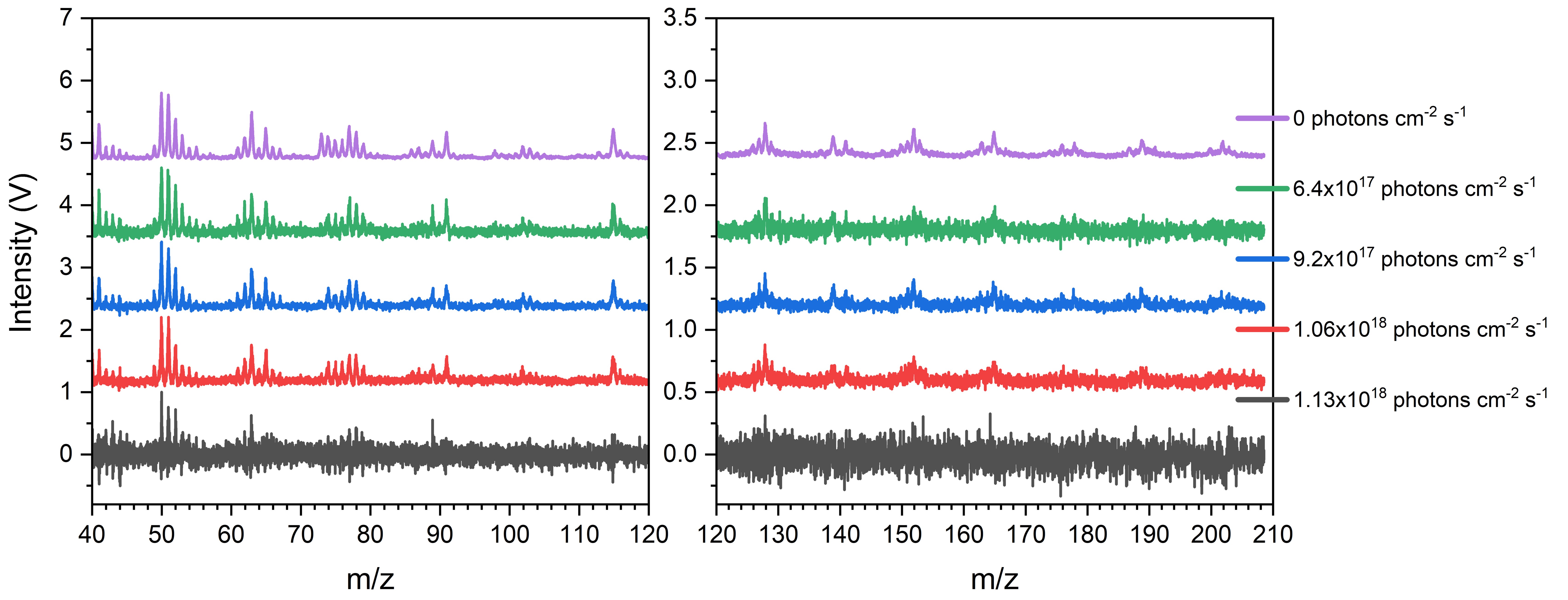}
    \caption{LDPI ReTOF-MS of the organic residue produced upon VUV irradiation of C$_2$H$_2$ ice at 15~K, exposed to further VUV fluence is shown. The fluence in the legend indicates the total photon fluence the formed residue was exposed to after the ice over a specific column was ablated, effectively halting the residue formation over that column. All the mass spectra were normalized to m/z = 50, for comparison.}
     \label{fig:fig006}
\end{figure*}

On the other hand, all the aforementioned m/z values corresponding to aromatic ions at m/z = 63, 115, 128, 165, 178, and 202, can be obtained by the sum of identified aliphatic ions, shown in Table~2. Indeed, the m/z = 77 fragment assigned to the phenyl group may correspond to contributions from benzene (C$_6$H$_6$) derivatives or various hexadiyne or hexadienyne groups, such as R-CH$_2$-C$\equiv$C-CH$_2$-C$\equiv$CH, R-C$\equiv$C-CH$_2$-C$\equiv$C-CH$_3$, R-C$\equiv$C-CH$_2$-CH$_2$-C$\equiv$CH (all are m/z = 39 + 38 = 77) or R-CH=CH-CH=CH–C$\equiv$CH, R-C$\equiv$C-CH=CH-CH=CH$_2$, R-CH=CH-C$\equiv$C-CH=CH$_2$ (all are m/z = 50 (51) + 27 (26) (deflected)= 77). Similarly, the m/z = 128 value assigned to naphthalene can be obtained by various substituted benzol fragments or even non-aromatic polyyne fragments. It should be noted that application of LDPI ReTOF-MS does not allow for the distinction between the cumulenic molecules (molecules containing 3 or more consecutive double bonds between C atoms) and polyynic structures \citep{Cernicharo_Agundez_Cabezas_Tercero_Marcelino_Pardo_de_Vicente_2021}. The preference for the assignment of polyynes and polyenynes is given due to the higher stability of these isomers and the high excess energy of VUV-photon absorption-induced reactions involved in the experiments. Despite the many aliphatic and aromatic candidates for our mass spectra, the exact aliphatic to aromatic product ratios cannot be determined using our experimental technique. 

The preference for the partial assignment of aromatic fragments is supported by the identification of benzol, phenylacetylene, styrene, naphthalene, and phenanthrene among the volatile products of C$_2$H$_2$ ice radiolysis by \cite{Abplanalp_Kaiser_2020} and the consistency of their results with this work, shown in section 3.2. \cite{Abplanalp_Kaiser_2020} reports a relatively large span of possible ratios between the primary aliphatic product vinylacetylene and the primary aromatic product benzene, ranging from 8 to 0.5, depending on the applied detection technique, shown in Tables 5 and 6 of \cite{Abplanalp_Kaiser_2020}. This converges into the ratio from 5 to 0.1 for the number of carbon atoms locked in the vinylacetylene and benzene, respectively. The reported abundances of substituted arenes and large PAHs are at least a factor of a few lower than those of benzene. Unfortunately, \cite{Abplanalp_Kaiser_2020} does not report the abundances of 6-carbon-bearing aliphatic species, which is likely caused by the limitation of the applied techniques. Nevertheless, the composition of the produced organic residues should reflect the volatile inventory produced upon irradiation of C$_2$H$_2$ ice. Therefore, the presence of aromatic fragments with ratios similar to those reported by \cite{Abplanalp_Kaiser_2020} may be expected for our obtained residues.

The photostability of the produced organic residue upon VUV-irradiation is measured in this work as a function of VUV fluence. Fig.~5 presents the mass spectra of five different organic residues exposed to five different VUV irradiation doses. All five residues are obtained on the same substrate by irradiation of 100~ML thick pure C$_2$H$_2$ ice in the following way. The first residue is obtained by ablating volatile components from a first column of the ice, after a total of 4 minutes ($\sim$7 $\times$ 10$^{16}$~photons~cm$^{-2}$ or $\sim$7~eV~per~molecule) of irradiation. Similarly, a second residue is obtained by ablating volatile ice components from a second fresh column after 8 minutes ($\sim$1.4 $\times$ 10$^{17}$~photons~cm$^{-2}$ or $\sim$14~eV~per~molecule) of irradiation. The procedure is repeated until the ablation of the last column after 64 minutes (1.2 $\times$ 10$^{18}$~photons~cm$^{-2}$ or $\sim$120~eV~per~molecule) of irradiation. Since the substrate, including the produced residues, was continuously exposed to VUV-photon fluence over the entire duration of the experiment, the residue produced after 4 minutes of irradiation was exposed to a total of $\sim$1.13 $\times$ 10$^{18}$~photons~cm$^{-2}$ of VUV exposure. Similarly, the residue obtained after a total of 8 minutes of irradiation was exposed to a total of $\sim$1.06 $\times$ 10$^{18}$~photons~cm$^{-2}$ of total VUV exposure, and so on. The LDPI ReTOF mass spectra of the residues are collected at 15~K following an annealing overnight at 300~K. The obtained mass spectra were normalized to the intensity of mass m/z = 50 for easier relative comparison. All acquired residue mass spectra presented in Fig.~5 are qualitatively identical except for a weak signal appearing on the final spectrum at m/z = 73. This indicates that the formation of organic residues occurs very early during the irradiation process upon receiving several eV (1 to 7)~per~molecule irradiation dose. Moreover, the produced organic residue remains qualitatively unchanged upon prolonged irradiation by about 1$\times$10$^{18}$ of total VUV photon fluence.

\section{\label{sec:level4} Astrophysical implications and conclusions \protect}

In this experimental work, we confirm the efficient formation of several low-volatile hydrocarbons, including polycyclic aromatic hydrocarbons (PAHs), by investigating the UV photolysis of pure C$_2$H$_2$ ices at 15~K. The experimental results are in line with previous works using 5~keV electron bombardment of C$_2$H$_2$ ice; shown in Table~1. We also characterize several ML thick organic refractory residues produced by irradiation of C$_2$H$_2$ ice under ultra-high vacuum conditions. The organic residues are, for the first time, characterized in situ by LDPI ReTOF-MS after annealing the processed ice sample at 300~K. This approach preserves the pristine chemical composition upon VUV irradiation by avoiding chemical alteration caused by exposure to atmospheric conditions.

PAH formation in the ISM has drawn considerable attention over the past decades and has been proposed as a potential molecular carrier for the unidentified infrared emission (UIE) bands \citep{Li_Greenberg_1997, Tielens_2008, Kwok_Zhang_2011}. This attention has increased by both the recent detection of the first individual aromatic molecules in the ISM \citep{McGuire_Burkhardt_Kalenskii_Shingledecker_Remijan_Herbst_McCarthy_2018, McGuire_Loomis_Burkhardt_Lee_Shingledecker_Charnley_Cooke_Cordiner_Herbst_Kalenskii_etal._2021, Wenzel_Speak_Changala_Willis_Burkhardt_Zhang_Bergin_Byrne_Charnley_Fried_etal._2025, Cernicharo_Agundez_Cabezas_Tercero_Marcelino_Pardo_de_Vicente_2021, Burkhardt_Long_Kelvin_Lee_Bryan_Changala_Shingledecker_Cooke_Loomis_Wei_Charnley_Herbst_McCarthy_etal_2021} and by the high predicted abundances of PAHs in the ISM. PAHs have been suggested to contribute to dust grain formation, VUV shielding in star-forming regions, catalytic properties, and the formation of refractory organic residues. They have been suggested to be important precursors of many complex organic molecules, including prebiotic molecules \citep{Greenberg_Gillette_Caro_Mahajan_Zare_Li_Schutte_Groot_Mendoza-Gómez_2000, Ehrenfreund_Cami_2010, Accolla_Pellegrino_Baratta_Condorelli_Fedoseev_Scirè_Palumbo_Strazzulla_2018}. The formation of PAHs is generally explained via a top-down or bottom-up formation mechanism, depending on the environment \citep{Reizer_Viskolcz_Fiser_2022, Concepcion_Jimenez-Serra_Rivilla_Colzi_Martin-Pintado_2023}. In the top-down mechanism, the formation of PAHs is attributed to the fragmentation or destruction of larger carbon allotropes such as graphite, graphene, large PAHs, and carbon nanoparticles \citep{Merino_Svec_Martinez_Jelinek_Lacovig_Dalmiglio_Lizzit_Soukiassian_Cernicharo_Martin-Gago_2014, Boersma_Bregman_Allamandola_2013, Boersma_Bregman_Allamandola_2015}. On the other hand, the bottom-up mechanism proposes that PAHs can be formed in the ISM from simple hydrocarbons. Gaseous C$_2$H$_2$, abundant in the outflows of carbon-rich stars together with its derivatives C$_2$H, serves as one of the possible building blocks for PAH synthesis upon UV photolysis or radiolysis by energetic particles \citep{Abplanalp_Kaiser_2020, Zhang_Wang_Turner_Marks_Chandra_Fortenberry_Kaiser_2023, Pentsak_Murga_Ananikov_2024, Yang_Troy_Xu_Kostko_Ahmed_Mebel_Kaiser_2016}.

\begin{figure*}[htp]
   \centering
     \includegraphics[scale=0.7]{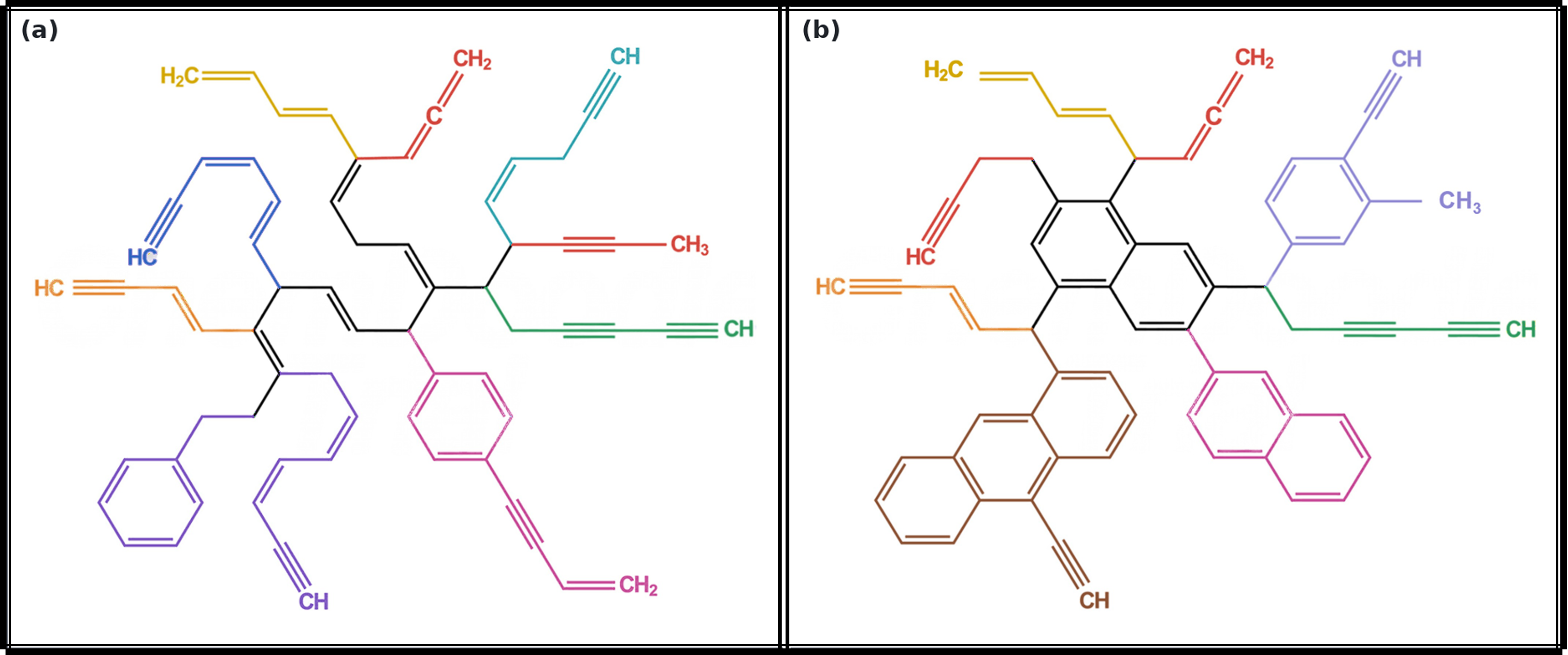}
    \caption{Two proposed residual structures originating from the UV photolysis of C$_2$H$_2$ ice: (a) low degrees of aromaticity and (b) high degrees of aromaticity. Some of the dissociative ionization fragments are shown in different colors. These substituted functional groups are: m/z = 39 (red), m/z = 51 (orange), m/z = 53 (yellow), m/z = 63 (green), m/z = 65 (turquoise), m/z = 77 (blue), m/z = 91 (violet), m/z = 115 (lavender), m/z = 127 (pink), m/z = 201 (brown). The molecules are presented in two-dimensional flat structures for simplicity.}
     \label{fig:fig007}
\end{figure*}

In this work, the formation of large unsaturated hydrocarbons containing up to 13 carbon atoms is found in the VUV-processed C$_2$H$_2$ with a fluence of approximately 1–10$ \times 10^{17}$~photons~cm$^{-2}$. This corresponds to a timescale of roughly $3 \times 10^{5}$–$3 \times 10^{6}$ years in dense molecular clouds with a UV flux of about $10^{4}$~photons~cm$^{-2}$~s$^{-1}$ \citep{Prasad_Tarafdar_1983}. The identified PAH products align with the outcome of the similar work by \citet{Abplanalp_Kaiser_2020} using REMPI ReTOF-MS following bombardment of pure (C$_2$H$_2$) ice by 5~keV electrons at 5~K. All the characteristic mass signals corresponding to benzene (C$_6$H$_6$), phenylacetylene (C$_8$H$_6$), styrene (C$_8$H$_8$), naphthalene (C$_{10}$H$_8$), and phenanthrene (C$_{14}$H$_{10}$) are securely observed in the LDPI ReTOF mass spectra obtained at 15~K without warming up the processed ice samples. However, some internal isomerization of unsaturated aliphatic hydrocarbon ions into more stable aromatic formats upon electron impact cannot be completely excluded. For the same reason, the degree of aromaticity among the photolysis products cannot be derived. \citet{Abplanalp_Kaiser_2020} demonstrate a wide range of possible ratios between the number of carbon atoms incorporated in the two main polymerization products, aliphatic and aromatic, ranging from 5 to 0.1. Nevertheless, their results clearly hint at the mixed aliphatic-aromatic nature of the obtained radiolysis products. Similarly, the refractory organic residues produced by our UV photolysis of pure C$_2$H$_2$ ice at 15~K are inferred to have mixed aliphatic-aromatic composition. By carefully analyzing the dissociative mass fragments, the residues formed in our experiments are predominantly composed of conjugated unsaturated triple bonds (-C$\equiv$C-) and double bonds (>C=C<). Moreover, the obtained mass spectra are dominated by mass signals characteristic of aromatic dissociative ionization fragments, suggesting the presence of aromatic structures containing one to four fused rings (see Table~2 for details).

Based on the qualitative data from the residual compound analysis, two possible refractory structures are proposed in Fig.~6. Fig.~6(a) corresponds to the structure featuring aliphatic carbon chains with a low degree of aromaticity. Various functional groups attached to the base carbonaceous compound, as evidenced by the detection of mass signals listed in Table~2, are highlighted by different colors. Alternatively, Fig.~6(b) depicts the refractory structure which is rich in (fused) aromatic rings. Likewise, the substituted functional groups responsible for the observed signals are highlighted in different colors. It should be noted that both proposed residual structures demonstrate a good match with possible carriers for the unidentified infrared emission features illustrated in \citet{Yang_Li_Glaser_Zhong_2017}. These are interstellar hydrogenated amorphous carbon (HAC) \citep{Duley_Jones_Williams_1989, Jones_Duley_Williams_1990}, and to some extent, quenched carbonaceous composite (QCC) \citep{Sakata_Wada_Onaka_Tokunaga_1990} and mixed aromatic-aliphatic organic nanoparticles (MAON) structures. The latter two structures contain O and N atoms, which are missing in pure C$_2$H$_2$ photolysis experiments. Irradiation of C$_2$H$_2$ with inclusions of O- and N-bearing species can be the subject of future studies.

In the UV photolysis experiments presented in this work, the organic residues are synthesized at the beginning of VUV exposure with a fluence of 7 $\times$ 10$^{16}$~photons~cm$^{-2}$ and remain chemically stable with some changes in intensity, as seen in Fig.~5. This result suggests a strong photostability of the produced organic residues, which could likely be preserved on interstellar dust grains during stellar evolution. This is in line with the initial hypothesis that photostable refractory organic residues are produced on the surface of interstellar dust by energetic processing of icy mantles \citep{Pendleton_Allamandola_2002, Caro_Schutte_2003, Ehrenfreund_Charnley_2000, Kwok_2004}. Another important result is the acquisition of nm thick organic residues compared to previous experimental studies where residue formation was mainly observed after energetic processing of micrometer-thick volatile ices, such as carrying mixtures of CO:NH$_3$:H$_2$O:CH$_3$OH:CO$_2$ \citep{Greenberg_Gillette_MuñozCaro_Mahajan_Zare_Li_Schutte_deGroot_Mendoza-Gómez_2000, Ehrenfreund_B_Sephton_2006, Accolla_Pellegrino_Baratta_Condorelli_Fedoseev_Scirè_Palumbo_Strazzulla_2018}. In this work, VUV photolysis of 100~ML thick ($\sim$30 nm) ice, representative of observational ice thicknesses in the ISM \citep{Boogert_Gerakines_Whittet_2015}, leads to
abundant detection of organic residue within the typical molecular cloud timescale of 10$^6$ years. Modeling results obtained for the outflows of carbon-rich AGB stars estimate ice thicknesses which are of a fraction of ML in the outer cold part of the circumstellar envelope (Vande Sande et al.\ \citeyear{VandeSande_Walsh_Mangan_Decin_2019}, \citeyear{VandeSande_Walsh_Danilovich_2020}, \citeyear{VandeSande_Walsh_Millar_2021}).

This study serves as a proof-of-principle experiment, demonstrating that refractory organic residues with astrophysical significance can be synthesized from astronomically thin ices and characterized under carefully controlled UHV conditions, without external contaminants. This study also provides insights into what the inner refractory mantles would look like on dust grains. Future experiments will extend these techniques to more astrophysically realistic ice mixtures, including multiple volatile species relevant to dense molecular clouds. Additionally, complementary techniques such as FTIR or Raman spectroscopy can be used to characterize the residues and nonvolatile photoproducts. Such follow-up studies will provide deeper insights into the complexity of refractory materials produced in diverse astrophysical environments and their potential roles in prebiotic chemistry.

 \begin{acknowledgements}

This work has been supported by the Danish National Research Foundation through the Center of Excellence InterCat (Grant Agreement No. DNRF150). It has also been funded by the Dutch Astrochemistry Network II (DANII) and NOVA (the Netherlands Research School for Astronomy). GF acknowledges the Xinjiang Tianchi Talent Program (2024).

\end{acknowledgements}

\bibliographystyle{aa} 
\bibliography{Acetylene}

\begin{appendix} %First appendix
\section{Acetylene cluster ions versus hydrocarbon impurities}

Fig.~A.1 presents three LDPI ReTOF mass spectra obtained in a set of control experiments aimed at confirming that the mass features found in our pure  unirradiated C$_2$H$_2$ ice mass spectrum obtained at 15~K (see Fig.~2) above m/z = 30 were indeed caused by C$_2$H$_2$ clusters rather than contaminants. The red spectrum is obtained for 100~ML thick C$_2$H$_2$ ice after deposition at 15~K and warm up to 60~K. The blue spectrum is obtained after a similar deposition and warm-up to 80~K. The 60~K and 80~K substrate temperatures lie before and after the thermal desorption temperature of pure C$_2$H$_2$ ice at 78~K under our applied UHV conditions. The mass spectrum at 15~K before the deposition of the pure C$_2$H$_2$ ice is presented for comparison. To ensure reliable comparison between spectra and reduce experimental uncertainties, all measurements were conducted using the same instrumental settings except for the substrate temperature. The spectrum presented in Fig.~A.1 reveals an absence of any hydrocarbon mass signals for both the pre-deposition mass spectra obtained at 15~K (black trace) and post-deposition mass spectra of C$_2$H$_2$ ice heated to 80~K (blue trace). The mass spectrum at 80~K contains trace signals of H$_2$O ice and leftover C$_2$H$_2$ at masses m/z = 18 and 26, respectively. The mass spectrum obtained after the C$_2$H$_2$ ice is warmed up to 60~K (red) shows identical mass features to the mass spectrum of the pure C$_2$H$_2$ ice obtained at 15~K, shown in Fig.~2. This confirms that the peaks seen in the mass spectra of C$_2$H$_2$ ice at 15~K, especially the peaks at higher masses, must come from C$_2$H$_2$ ice clustering upon laser ablation, rather than contaminants from the gas bottle or interactions with the ion optics of the setup.

\begin{figure}[htp]
   \centering
     \includegraphics[width = 0.8 \linewidth]{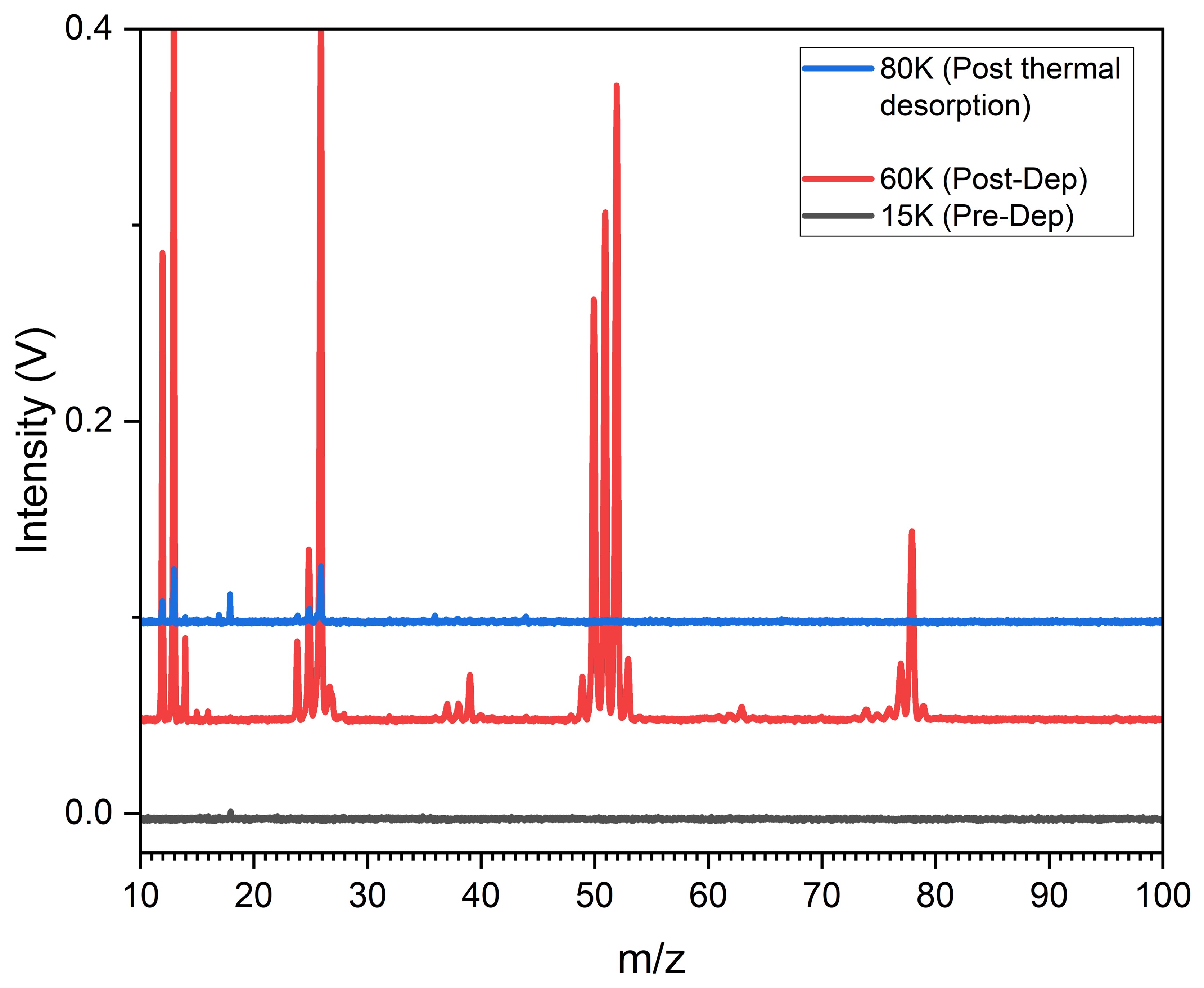}
    \caption{Comparison of the LDPI ReTOF mass spectrum of a clean substrate, obtained before the deposition of C$_2$H$_2$ ice at 15~K (black), a spectrum of 100~ML thick C$_2$H$_2$ ice obtained after the deposition at 15~K and warmed up to 60~K (red), and the spectrum obtained after annealing of the same C$_2$H$_2$ ice at 80~K (blue). The thermal desorption temperature of pure C$_2$H$_2$ ice under our applied UHV conditions is about 78~K. The presence of hydrocarbon features in the mass spectrum obtained at 60~K and their absence in the spectrum collected at 80~K is consistent with the assignment of signals above m/z = 28 to the acetylene cluster ions (C$_2$H$_2$)$_n$$^+$, where n = 2, 3, ..., 7.}
     \label{fig:fig008}
\end{figure}

\section{Excluding large saturated aliphatic hydrocarbons}

Fig.~B.1 presents a comparison between the mass spectrum of an unsaturated aromatic hydrocarbon, naphthalene (C$_{10}$H$_8$) and a fully saturated aliphatic hydrocarbon, nonane (C$_9$H$_{20}$). These two hydrocarbons are isobaric, each exhibiting a molecular mass of 128 amu. The mass spectrum of nonane is dominated by signals in the lower mass region, especially around m/z = 35 to m/z = 65. In contrast, the mass spectrum of naphthalene is dominated by a few m/z signals around the parent ion m/z = 128. The LDPI ReTOF mass spectra obtained after VUV-irradiation of pure C$_2$H$_2$ ice are comprised of several narrow groups of m/z signals in the higher mass range (m/z > 75), which are more consistent with the presence of various unsaturated (aromatic) hydrocarbons.

\begin{figure}[htp]
   \centering
     \includegraphics[width = 0.7\linewidth]{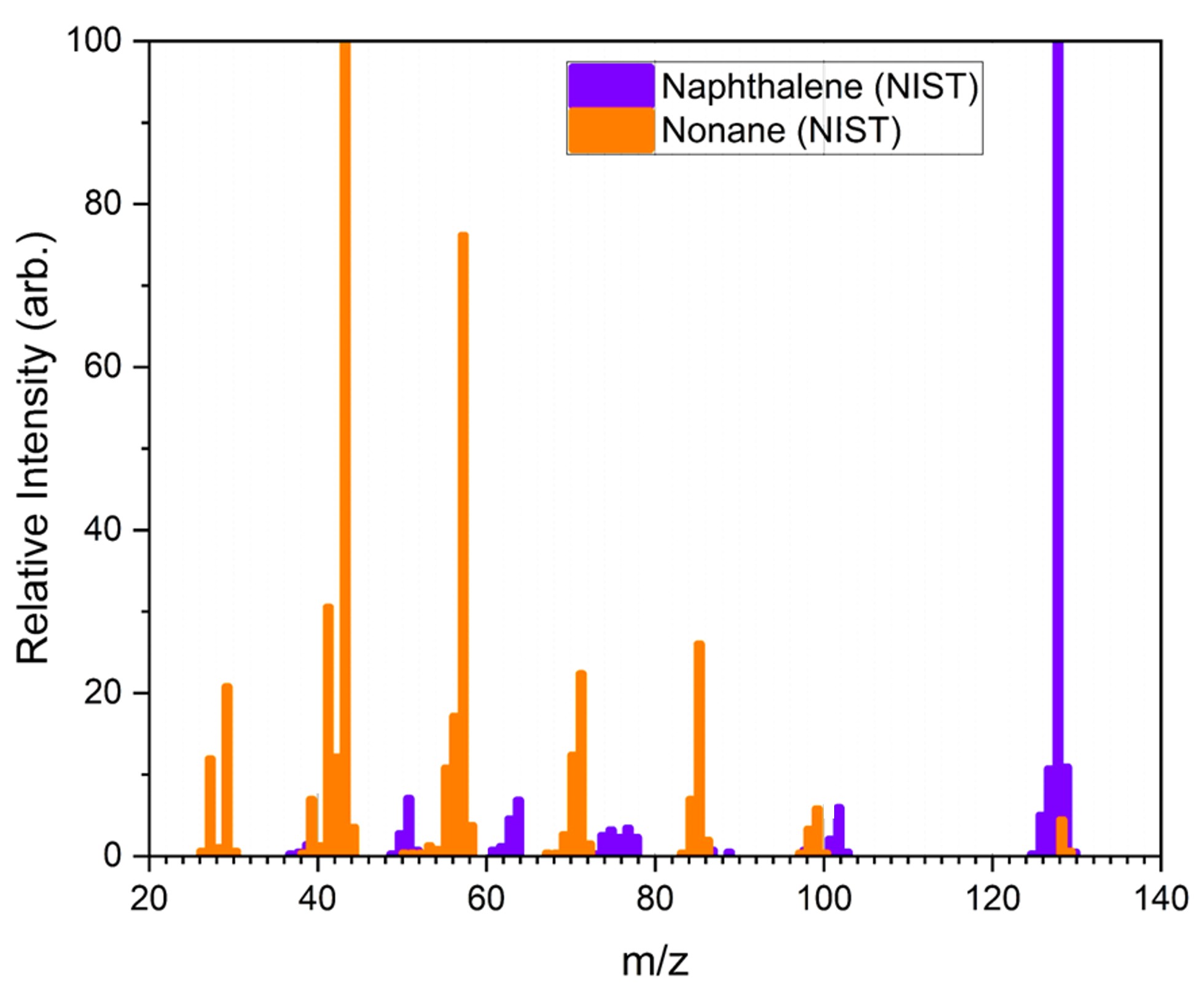}
    \caption{Comparison between the mass spectra of two isobaric hydrocarbons with m/z = 128, naphthalene (C$_{10}$H$_8$, in purple), and nonane (C$_9$H$_{20}$, in orange). The mass spectra are obtained from the National Institute of Standards and Technology (NIST) database \citep{NIST_ChemWebBook}.}
     \label{fig:fig009}
\end{figure}

\section{Uniformity of the residue}

Fig.~C.1 presents the LDPI ReTOF mass spectra acquired from five distinct columns on the substrate. The spectra were collected for the residue formed after the VUV irradiation of pure C$_2$H$_2$ ice with a fluence of 6 $\times$ 10$^{17}$~photons~cm$^{-2}$ (32 minutes of irradiation). The mass spectra were collected following overnight annealing of the substrate at 300~K and subsequent cooling to 15~K, in order to maximize the thermal desorption of volatiles and minimize the contamination from residual chamber gases. The mass features at m/z = 17 and m/z = 18 arise from the water H$_2$O ice deposited on the substrate upon cooling down to 15~K. To highlight the uniformity of the mass spectral features of the produced residue and their relative intensities across the substrate, the spectra are presented as collected, without normalization. The mass spectra recorded across the five sampled columns on the substrate exhibit qualitatively identical mass features, with only minor fluctuations in peak intensities varying from column to column. This high degree of spectral consistency suggests that the chemical composition of the residue is largely uniform across the substrate.

\begin{figure*}[htp]
   \centering
     \includegraphics[width = 1.25\linewidth]{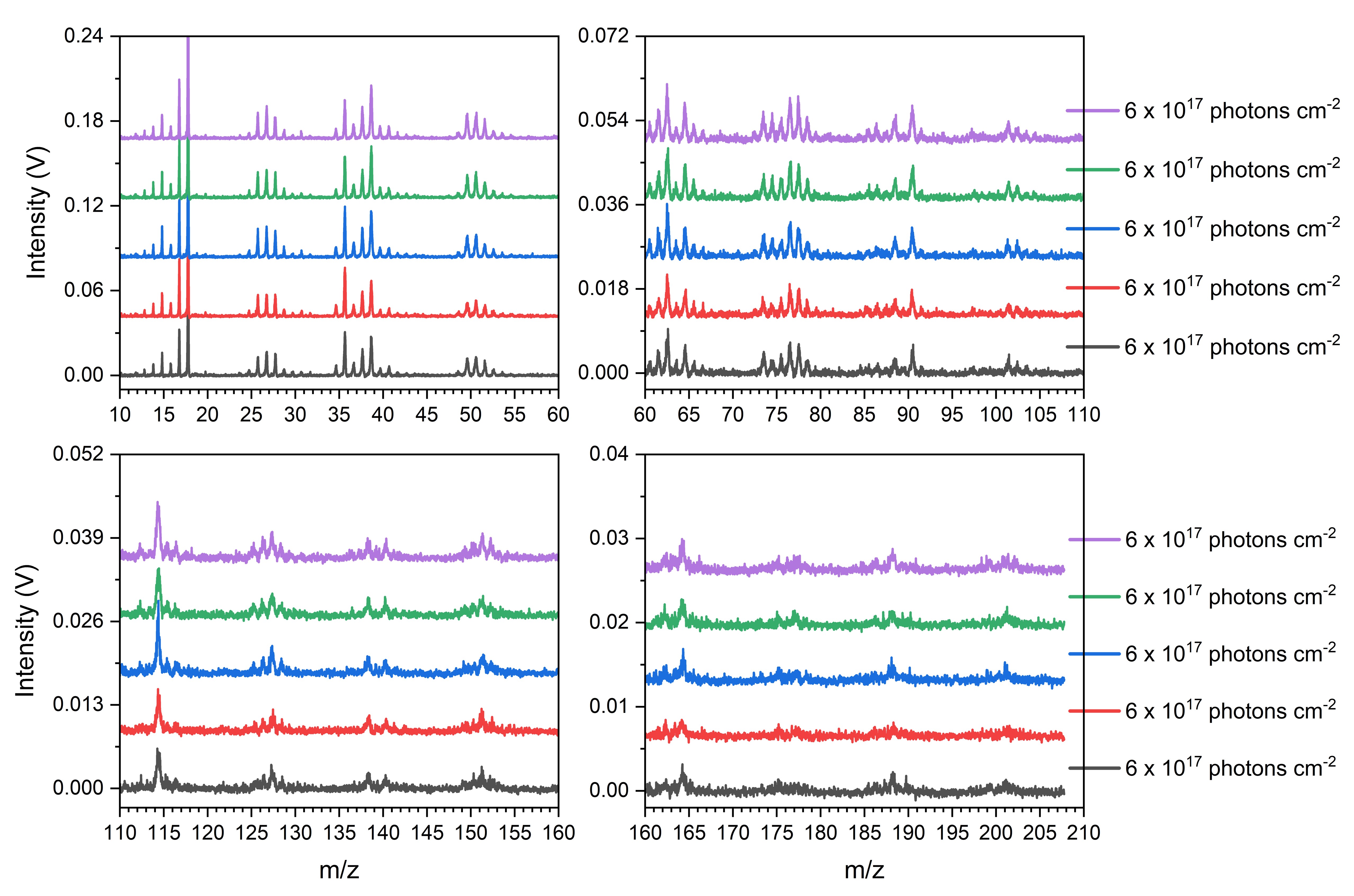}
    \caption{Comparison of the LDPI ReTOF mass spectra of the residue produced after VUV irradiating C$_2$H$_2$ ice with 6 $\times$ 10$^{17}$~photons~cm$^{-2}$. The mass spectra, obtained at 15~K after annealing overnight at 300~K, are shown across multiple laser-probed columns on the substrate. This figure showcases the uniformity of peaks and consistency of the formed residue across 5 columns probed on the substrate.}
     \label{fig:fig010}
\end{figure*}
\end{appendix}

\end{document}